\documentclass[letterpaper,12pt]{article}

\usepackage[linkcolor=blue,urlcolor=blue,citecolor=blue, colorlinks,linktocpage,bookmarks, pdfstartview=FitH]{hyperref}
\usepackage{natbib}
\usepackage{graphicx}
\usepackage{fancyhdr}
\usepackage[all]{hypcap}
\usepackage{latexsym}
\usepackage{amsmath}
\usepackage{verbatim}
\usepackage{setspace}
\usepackage{subfig}
\usepackage{titlesec}
\usepackage{enumitem}

\allowdisplaybreaks
\makeatletter
\g@addto@macro\normalsize{
\setlength\abovedisplayskip{5pt}
\setlength\belowdisplayskip{5pt}
}
\makeatother

\titlespacing*{\section}{0pt}{-5pt}{-5pt}
\titlespacing*{\subsection}{0pt}{-10pt}{-5pt}

\begin{document}

\bigskip

\begin{center}

\doublespacing

~

{\Large \bf Fully Bayesian Logistic Regression with Hyper-LASSO Priors for High-dimensional Feature Selection}

 \bigskip

Longhai Li\footnotemark[1] and Weixin Yao\footnotemark[2]

\smallskip
  
 \today 
%12 May 2014
\end{center}

% !TEX root =  bplrpaper.tex
\onehalfspacing
%\doublespacing

\begin{center}
\large\textbf{Abstract}
\end{center}
\noindent High-dimensional feature selection arises in many areas of modern science.  For example, in genomic research we want to find the genes that can be used to separate tissues of different classes (\textit{e.g.} cancer and normal) from tens of thousands of genes that are active (expressed) in certain tissue cells.  To this end, we wish to fit regression and classification models with a large number of features (also called variables, predictors).  In the past decade,  penalized likelihood methods for fitting regression models based on hyper-LASSO penalization have received increasing attention in the literature. However, fully Bayesian methods that use Markov chain Monte Carlo (MCMC) are still in lack of development in the literature.  In this paper we introduce an MCMC (fully Bayesian) method for learning severely multi-modal posteriors of logistic regression models based on hyper-LASSO priors (non-convex penalties). Our MCMC algorithm uses Hamiltonian Monte Carlo in a restricted Gibbs sampling framework; we call our method Bayesian logistic regression with hyper-LASSO (BLRHL) priors.  We have used simulation studies and real data analysis to demonstrate the superior performance of hyper-LASSO priors, and to investigate the issues of choosing heaviness and scale of hyper-LASSO priors. %  An R add-on package called \texttt{BLRHL} will be available from  \url{http://math.usask.ca/longhai/} and CRAN.

\noindent \textbf{Key phrases:} high-dimensional, feature selection, non-convex penalties, horseshoe, heavy-tailed prior, hyper-LASSO priors, MCMC, Hamiltonian Monte Carlo, Gibbs sampling, fully Bayesian.

\vfill

\footnotetext[1]{\onehalfspacing Correspondence author; Department of Mathematics and Statistics, University of Saskatchewan, Saskatoon, SK, S7N5E6, CANADA. e-mail: \texttt{longhai@math.usask.ca}. The research of Longhai Li was supported by the fundings from Natural Sciences and Engineering Research Council of Canada (NSERC), and Canada Foundation of Innovations (CFI). }

\footnotetext[2]{\onehalfspacing Department of Statistics, University of California at Riverside,  Riverside, CA, 92521, USA. Email: \texttt{weixin.yao@ucr.edu}. The research of Weixin Yao was supported by NSF grant DMS-1461677.}

\newpage

% !TEX root =  bplrpaper.tex

\def\mb#1{\mbox{\boldmath $#1$}}	  
\def\IG{\mbox{IG}}
\def\G{\mbox{Gamma}}
\def\hblr{\textbf{BLRHL} }

\def\mb#1{\mbox{\boldmath $#1$}}
\def\IID{\,\begin{array}{cc} \\[-20pt] \mbox{\tiny IID} \\[-8pt] \sim
\end{array}\,}
\def \given{\,|\,}
\def\mathscript#1{\hbox{\tiny$#1$}}

\def\Vj {V(\mb\delta_{j,1:K})}
\def\tp {\textit{t} prior } 
\def\df {\mbox{df}}

\doublespacing

\section{Introduction} \label{sec:intro}

The accelerated development of many high-throughput biotechnologies has made it affordable to collect measurements of high-dimensional molecular changes in cells, such as expressions of genes. These gene expressions are referred to as \textit{features} generally in this paper, and are often called \textit{signatures} in life sciences literature. Scientists are interested in selecting features related to a categorical response variable, such as cancer onset or progression.     Identifying the most relevant genes for a disease from a large number of candidates is still a tremendous challenge to date; an analogy is that we are looking for a few ``needles'' (useful features) from a huge ``haystack'' (unrelated features).

The most widely used methods in today's practice are univariate methods that measure the strength of the relationship between each gene and the class label, \textit{e.g.} $t$ or $F$ tests, or model-based inference methods where independence is assumed for genes within classes, \textit{e.g.} DLDA \citep{dudoit2002comparison}, and PAM \citep{tibshirani2002dmc}. A major issue with univariate methods is that they ignore the correlations between genes which are prevalent in gene expression data due to gene co-regulation, see \citet{ma2007supervised}, \citet{clarke2008properties} and \cite{tolosilengauer2011} for real examples. One consequence of this ignorance is that  many redundant differentiated genes are included, while useful but weakly differentiated genes may be omitted. 

Methods of fitting classification/regression models which attempt to capture the conditional distribution of class label (response) given features can take correlations among features into account. However, when the number of observations is not much larger than the number of features, maximizing likelihood of a classification model will overfit the data, with noise rather than signal being captured.  Therefore, when the number of features is greater than the number of observations, we need to shrink the coefficients toward 0 to avoid overfitting. The most widely used method to achieve this is LASSO \citep{tibshirani1996regression}, which uses a {\em convex} $L_1$ penalty (or Laplace prior in Bayesian inference); however, Laplace prior cannot effectively distinguish the ``needles'' and ``hay'' due to its light tails. Considering the super-sparsity of important features related to a response, many researchers have proposed to fit classification or regression models using continuous non-convex penalty functions.   This approach, which has been given the names hyper-LASSO or global-local penalities, has been widely recognized for its ability to shrink the coefficients of unrelated  features (noise) more aggressively to 0 than LASSO while retaining the significantly large coefficients (signal). In other words, non-convex penalties provide a sharper separation of signal from noise. Such non-convex penalties include (but are not limited to): a $t$ distribution with a small degree of freedom \citep{gelman2008weakly,yi2012hierarchical}, SCAD \citep{fan2001variable}, horseshoe  \citep{gelman2006prior, carvalho2009handling, carvalho2010horseshoe, polson2012half-cauchy,van_der_pas2014horseshoe},  MCP \citep{zhang2010nearly}, NEG \citep{griffin2011bayesian},   adaptive LASSO \citep{zou2006adaptive}, generalized double-pareto \citep{armagan2010bayesian}, Dirichlet-Laplace and Dirichlet-Gaussian \citep{bhattacharya2012bayesian}, among others. For reviews of non-convex penalty functions, see \citet{kyung2010penalized,polson2010shrink,  polson2012good} and \citet{polson2012local};  \citet{breheny2011coordinate} and \citet{wang2014optimal} study the computational and convergence properties of optimization algorithms with non-convex penalization.  

Besides sparsity of signal, high-dimensional features often have a grouping structure or high correlation;  this often has a biological basis, for example  a group of genes may relate to the same molecular pathway, are in close proximity in the genome sequence, or share a similar methylation profile  \citep{clarke2008properties,tolosi2011classification}.  For such datasets, a non-convex penalty will make a selection within a group of highly correlated features; this will either split important features into different modes of penalized likelihood,  or suppress less important features in favour of more important features.  The within-group selection is indeed a desired property if our goal is to select a sparse subset of features; however, note that this within-group selection does not mean that we will lose other features within a group that are also related to the response because other features can still be identified from the group representatives using the correlation structure.  On the other hand, the within-group selection results in a very large number of modes in the posterior (for example, two groups of 100 features can make $100^{2}$ subsets containing one from each group). Because of this, optimization algorithms encounter great difficulty in reaching a global or good mode because in the non-convex region, solution paths are discontinuous and erratic. Although superior properties  of non-convex penalties (compared to LASSO) have been theoretically proved in statistics literature, many researchers and practitioners are still reluctant to embrace these methods due to their lack of convexity, because non-convex objective functions are difficult to optimize and often produce unstable solutions~\citep{breheny2011coordinate}.

The fully Bayesian approach---using Markov chain Monte Carlo (MCMC) methods to explore the multi-modal posterior---is a valuable alternative for non-convex learning, because a well-designed MCMC algorithm can travel across many modes. To the best of our knowledge there have been only a few reports in the literature discussing fully Bayesian (MCMC) methods for exploring regression posteriors based on hyper-LASSO priors; relevant articles include \citet{yi2012hierarchical}, \citet{piironen2016hyperprior}, \citet{nalenz2017tree} and possibly others.   In this paper we introduce an MCMC (fully Bayesian) method for learning severely multi-modal posteriors of logistic regression models based on hyper-LASSO priors (non-convex penalties). Our MCMC algorithm uses Hamiltonian Monte Carlo \citep{neal2010mcmc} in a restricted Gibbs sampling framework; this method substantially shortens the time it takes to sample high-dimensional but sparse regression coefficients.  The focus of this paper is placed on demonstrating the superior performance of hyper-LASSO priors, and investigating issues related to choosing the heaviness and scale of hyper-LASSO priors. Our empirical results show the following two properties of hyper-LASSO inference: first, the choice of degrees of freedom that control tail heaviness should be appropriate;  Cauchy appears optimal, which confirms the superior performance of horseshoe and NEG penalties in penalized likelihood methods.  Second,  due to the ``flatness'' in the tails of Cauchy,  the shrinkage of large coefficients is very small (\textit{i.e.}, small bias); more importantly, the shrinkage is very robust to the scale, which is a distinctive property of Cauchy priors compared to Laplace and Gaussian priors. 

This article will be structured as follows. In Section \ref{sec:simple}, we first discuss some properties of hyper-LASSO priors using comparisons to Laplace and Gaussian priors. In Section \ref{sec:meth}, we describe BLRHL in technical details. In Section \ref{sec:sim} we use simulated datasets to test our method and investigate the issue of choosing heaviness and scale. In Section \ref{sec:gene}, we report the results of our analysis by applying our methods to a real microarray dataset related to prostate cancer with $p=6033$. The article is concluded in Section \ref{sec:end} with discussions of future work.%; in particular, we will discuss how to interpret MCMC samples from many modes. 

\section{Properties of Hyper-LASSO Priors}\label{sec:simple}

We first consider the simple logistic regression model for binary class labels in order to examine two important properties of hyper-LASSO priors. Suppose we have collected data of features and responses (class labels) on $n$ training cases. For a case indexed by $i$, we denote its class label by $y_i$ (which can take integers $1 \mbox{ and } 2$), and denote the $p$ features associated with it by the row vector $\mb x_{i,1:p}$. The logistic regression model for the data is:
\begin{eqnarray}
P(y_i = k + 1| \mb x_{i,1:p}, \mb\beta_{0:p}) &=&
\dfrac{I(k = 0) + I(k = 1)\exp\left(\beta_{0} + \mb x_{i,1:p}
\mb\beta_{1:p}\right)}{1 + \exp\left(\beta_{0} + \mb x_{i,1:p}
\mb\beta_{1:p}\right)},
\label{eqn:binarylog}
\end{eqnarray}
for $k=0 \mbox{ and }1$, $i = 1,\ldots,n$, where $\mb\beta_{1:p}$ is a column
vector of regression coefficients and $I(\cdot)$ is the indicator function which
is equal to 1 if the condition in bracket is true, 0 otherwise. We will assume
that all features are commensurable and we will select features by looking
at the values of $\mb\beta_{1:p}$. We will consider how to infer
$\mb\beta_{1:p}$ from the training data. 

\iffalse
\textit{t} prior is very often used in low-dimensional
problems, see the textbook by \cite{bayesiandata}, and a recent article
\citep{Gelman08}. For high-dimensional genomic problems, however, using $t$
prior is rare, though not entirely absent. \citet{Bae04} used $t$ prior with
df=6 in probit  model for finding the useful genes from 500 genes that can
separate two classes of Leukemia. \cite{Yi08} applied regression models with
$t$ prior for finding QTLs from 127 gene markers, with dfs always greater than 1
and having a median 2.1 in MCMC sample. Both of these works did not see clear
difference of $t$ prior from Laplace, probably because their degree freedoms are
too large. We will show that small degree freedom and small scale for \textit{t}
prior is necessary for achieving clearly better results than LASSO in
high-dimensional problems. 
\fi

We consider scale-mixture-normal (SMN) distributions as priors for $\mb\beta_{1:p}$. The simplest choice for a prior is a 
$t$ distribution with $\alpha$ degrees of freedom and scale $\sqrt{w}$, which can be expressed with two-level conditional
distributions: $\beta_j | \sigma_j^2\sim N(0,\sigma_j^2), \sigma_j^2 \sim \IG
(\alpha/2, \alpha w/2)$, where $\IG(a,b)$ stands for Inverse-Gamma
distribution, the distribution of the inverse of a Gamma random variable with
shape parameter $a$ and \textit{rate} parameter $b$. In terms of a random number
generator, the products of two independent random
variables has a $t$-distribution with scale $\sqrt{w}$: $N(0,1)\times
\sqrt{\IG(\alpha/2,\alpha/2)}\times \sqrt{w}$. Similarly, 
$N(0,1)\times \sqrt{\exp (1)} \times \sqrt{w}$ has a Laplace distribution, where $\sqrt{w}$ is
the scale and $\exp (1)$ is the standard exponential random variable. Note that a
Laplace distribution parametrized by $\lambda$ with PDF $(\lambda/2) e^{-\lambda
|\beta_j|}$ has the scale $\sqrt{w} = \sqrt{2}/\lambda$. Recently, some other
penalties with SMN interpretation such as Horseshoe
\citep{carvalho2010horseshoe} and Normal-Exp-Gamma (NEG) 
\citep{griffin2011bayesian} have been shown to be superior than LASSO in
high-dimensional regression problems. In the original Horseshoe prior, a positive
half Cauchy prior is assigned to $\sigma_j$. Here we naturally generalize half
Cauchy to half \textit{t} for uniformity of notations for all the priors
considered in this article, and call the prior generalized horseshoe (GHS). In Table \ref{tab:smn}, we
describe them using their random number generators. The detailed descriptions of
GHS and NEG are given in Section \ref{sec:model} (Equations
\eqref{eqn:priorsgm-hs} and \eqref{eqn:priorsgmneg}). 

\begin{table}[t]
\centering
\caption{4 scale-mixture-normal distributions. }
\begin{tabular}{l|l}
\textbf{Name} & \textbf{Random Numbers Generator} \\ \hline
$t$     & $N(0,1)\times \sqrt{\IG(\alpha/2,\alpha/2)}\times \sqrt{w}$
         \\
GHS & $N(0,1)\times |N(0,1)|\times
            \sqrt{\IG (\alpha/2,\alpha/2)} \times \sqrt{w}$ 
           \\
NEG     & $N(0,1) \times \sqrt{\exp(1)}\times \sqrt{\IG(\alpha/2,\alpha/2)} 
          \times \sqrt{w}$ 
          \\
Laplace & $N(0,1) \times \sqrt{\exp (1)} \times \sqrt{w}$ 
       
\end{tabular}
\label{tab:smn}
\end{table}

\begin{figure}[htp]
\centering
\caption{Geometric illustrations of the properties of $t$ penalty in MAP inference.}
\subfloat [Distinguishing Needles from hay] {\label{fig:mappath}
\includegraphics[width=0.45\textwidth,height=0.27\textheight] {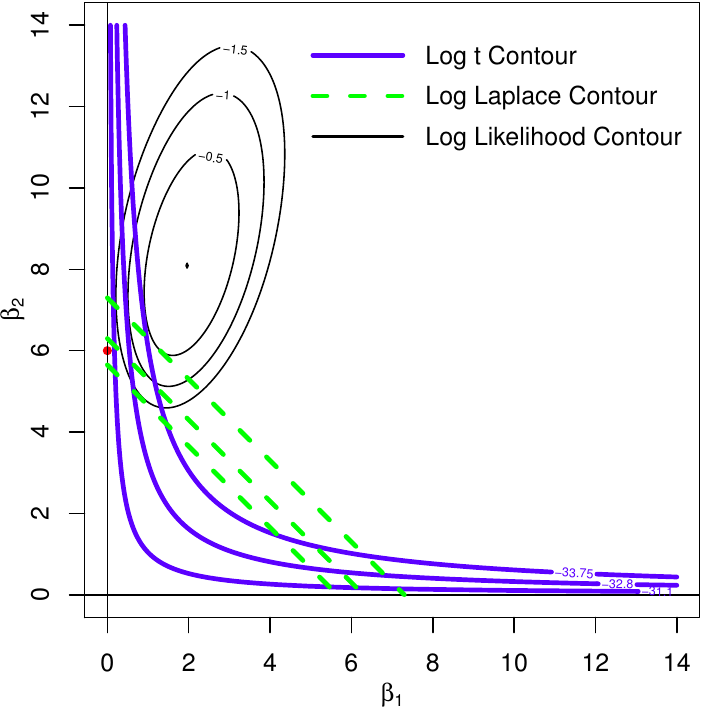} }
\subfloat [Dividing Redundant Correlated Features] {\label{fig:killcor}
\includegraphics[width=0.45\textwidth,height=0.27\textheight] {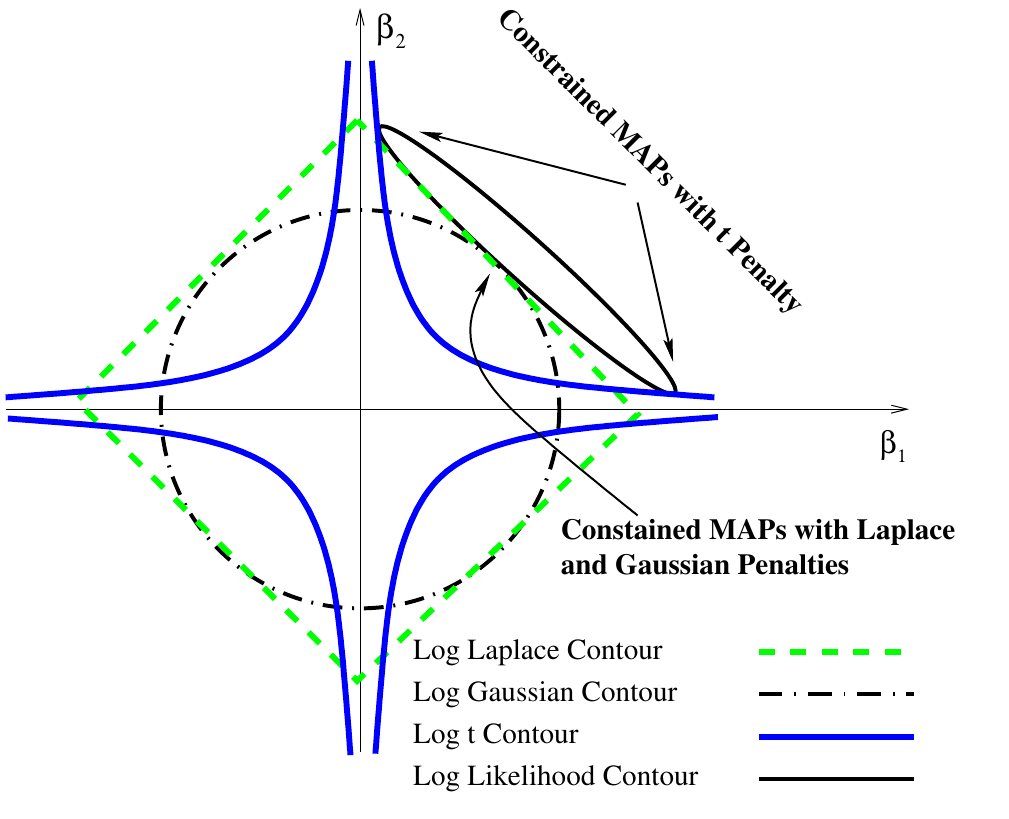} }
\label{fig:gtprior}
\vspace*{-20pt}
\end{figure}

The ability of moderately hyper-LASSO priors to better separate ``needles'' from ``hay'' can be explained by looking at the ``path'' of constrained MAPs (maximizer of posterior) ---  the MAP with the log likelihood constrained to a particular value (\textit{i.e.} on a contour). A constrained MAP can be found by shrinking the contour lines of log priors toward the origin until the two lines are tangent. Figure \ref{fig:mappath} shows three such constrained MAPs for a $t$ prior with df = 0.5 and  $\sqrt{w}=e^{-10}$, and three for a Laplace prior based on a dataset generated with true coefficients $\beta_1=0, \beta_2=6,\beta_0=0$. The (unconstrained) MAP can then be found from the ``path'' containing these constrained MAPs. Because the contour lines of the log $t$ prior indent into the origin, the path of constrained MAPs based on the $t$ prior is flatter (with respect to x-axis) than the path based on Laplace; starting from the MLEs, the path based on the $t$ prior goes to a point at which $\beta_1$ is very close to 0 (but not exact 0), and $\beta_2$ is close to its true value (6), whereas, the path based on Laplace goes to the origin. Therefore we see that the $t$ prior can shrink small ``hay'' without much punishment to large ``needles''.

From looking at constrained MAPs, we also find that hyper-LASSO penalties can \textit{automatically} divide a group of correlated features into different posterior local modes. Figure \ref{fig:killcor} shows a conceptual illustration. When two features are highly correlated, a contour line of log likelihood is negatively correlated as shown in Figure \ref{fig:killcor}. With a $t$ penalty, the constrained MAPs are at the two ends of the contour of log likelihood, each of which uses only one of them to explain the class label without underestimating the importance of each of them. Therefore, the coefficients of highly correlated features are divided into different modes, each using only one of them. When the predictive abilities of the correlated features are different, the \textit{t} prior can also make selections among a group of highly correlated features automatically. The selection within groups is necessary in high-dimensional problems in which  a large group of correlated features often exists. By contrast, with Laplace and Gaussian penalties, the constrained MAPs are in the middle of the contour, favoring using all features with coefficients of smaller absolute values to explain the class label. When the group of correlated features is large, they may underestimate the absolute values of all of them, and hence, miss all of them, see a detailed discussion by \cite{tolosilengauer2011}.

Figure \ref{fig:killcor} is also helpful for seeing that the primary
computational difficulty of using hyper-LASSO priors in classification and
regression problems is the presence of many local modes in the posterior
distribution. An optimization algorithm can easily get trapped in \textit{a}
minor local mode arbitrarily depending on the initial values, so the algorithm
becomes unstable and some sophisticated methods for choosing the initial values
are required, see \citet{griffin2011bayesian}. 

%Particularly, we choose sophisticated Hamiltonian Monte Carlo method for sampling the coefficients given hyperparameters. Our experiences show that this method can effectively travel across many modes, for which we make an explanation in Section \ref{sec:gibbs} with other computational details given.

\section{Bayesian Logistic Regression with Hyper-LASSO Priors} \label{sec:meth}

We will now describe our method, BLRHL, including some technical details. Throughout this article, we
will denote matrices with \textbf{bold-faced} letters, with row indexes
displayed in the first subscript and column indices in the second. We denote
real-valued vectors with \textbf{bold-faced} letters too, but with only a set of
indices in subscript. The indices of matrices and vectors are denoted by
$i\!:\!j$ --- integers from $i$ to $j$, or a single integer for a row or column.

\subsection{Multinomial Logistic Regression with Hyper-LASSO Priors} 
\label{sec:model}

Suppose we have collected data of features and responses (class labels) on $n$
training cases. For a case indexed by $i$, we denote its class label by $y_i$,
which can take integers $1,\ldots, C$, and denote $p$ features associated with
it by a row vector $\mb x_{i,1:p}$.  The hierarchical Bayesian multinomial
logistic regression model used by us is described as follows:
  \begin{eqnarray}
   P(y_i = c| \mb x_{i,1:p}, \mb\beta_{0:p,1:C}) &=&
   \dfrac{\exp\left(\beta_{0,c} + \mb x_{i,1:p}
\mb\beta_{1:p,c}\right)}{\sum_{c=1}^C\exp\left(\beta_{0,c} + \mb x_{i,1:p}
\mb\beta_{1:p,c}\right)}, \mbox{for }c = 1,\ldots, C,\label{eqn:ygivenx} \\[5pt]
   \mb \beta_{j,1:C} | \sigma_j^2 &\sim& N\ (0, \sigma_j^2), \ \mbox{for } j =
0,\ldots, p,
   \label{eqn:priorbeta}\\
   \sigma_{j}^2  &\sim& \mbox{IG}\ (\alpha/2, w\alpha/2), \mbox{for } j = 1,\ldots, p
\label{eqn:priorsgm}
   %\\
   %\log(w) &\sim& N(s, \eta^2), \label{eqn:priorw} 
  \end{eqnarray} 
where $\mb\beta_{0:p,1:C}$ are regression coefficients, and other variables are hyperparameters which are introduced to define the prior for
$\mb\beta_{1:p,1:C}$ (and for convenience in MCMC sampling). 

In this hierarchy, $\sigma_j^2$ indicates the importance of $j$th feature --- the feature with larger $\sigma_j^2$ is more useful for predicting $y$, provided that all features are commensurable (which can be enforced by standardization). Note that we fix $\sigma_0^2$, not controlled by $w$, because we believe that the variability of intercepts is quite different from the variability of $\mb\beta_{j,1:C}$ for features.  With $\mb\sigma^2_{1:p}$ marginalized with respect to IG prior \eqref{eqn:priorsgm}, Equations \eqref{eqn:priorbeta} and \eqref{eqn:priorsgm} assign $\mb\beta_{j,1:C}$ ($j>0$) a $C$-dimensional $t$ prior with $\alpha$ degrees of freedom and $I_C\times \sqrt{w}$ as its covariance, whose PDF can be found from \citet{kotz2004multivariate}. %For $C=2$, this bivariate $t$ prior is equivalent to assigning independent $t$ priors for  $p$ coefficients $\delta_{j} = \beta_{j,2}-\beta_{j,1}$.

An important issue in \textit{multinomial} logistic regression models is that the coefficients $\mb \beta_{j,1:C}$ are non-identifiable --- if we add a constant to all $\mb\beta_{j, 1:C}$, the conditional distribution of $y$ given $\mb x$ in \eqref{eqn:ygivenx} is \textit{exactly} the same. Therefore, the data can identify only the differences of $\mb\beta_{j,1:C}$ to a ``baseline'' class, say class 1, denoted by $\delta_{j,k}=\beta_{j,k+1} -\beta_{j,1}$, for $k=1,\ldots,C-1$.  To avoid the non-identifiability problem, the coefficient for a baseline class, say $\beta_{j, 1}$, is often fixed at 0, and $\mb\beta_{j,2:C}$ is assigned with a prior as in \eqref{eqn:priorbeta}. However, such a prior is \textit{asymmetric} for all classes: the prior variance of $\beta_{j,c}-\beta_{j,c'}$ ($c,c'\not=1$) double that of $\beta_{j,c} - \beta_{j,1}$. This implication may not be justified for practical problems. In addition, the feature selection can vary with the choice of baseline class.  %With fixed $\mb\delta_{j,1:(C-1)}$, for varying $\beta_{j,1}$, the normal prior [equation \eqref{eqn:priorbeta}] for $(\beta_{j,1}, \beta_{j,1}+\delta_{j,1},\ldots,\beta_{j,1} + \delta_{j,C-1})$ is maximized at $\beta_{j,1}=-(1/C)\sum_{k=1}^{C-1}\delta_{j,k}$, which implies $\sum_{c=1}^C\beta_{j,c}=0$. Therefore, the common variance prior, and so the posterior, identify the centralized $\mb\beta_{j,1:C}$.  However, the apparent non-identifiability in likelihood function does harm the efficiency of Markov chain sampling. A naive Gibbs sampler may stay for a long time where the absolute values of $\mb\beta_{j,1:C}$ are very large, but the differences $\mb\delta_{j,1:(C-1)}$ can be achieved by centralized $\mb\beta_{j,1:C}$ with smaller values. 

We can use symmetric and identifiable parameters in multinomial logistic regression by transferring the symmetric prior for $\beta$'s to the identifiable parameters $\delta$'s. The identifiable parameters for model \eqref{eqn:ygivenx} are defined as: 
\begin{eqnarray}
   \mb \delta_{j,k} &=& \mb \beta_{j,k + 1} - \mb \beta_{j, 1}, \ \ \mbox{for } k = 1, \ldots, \,\,K,\ \ j = 0,\ldots,p,  %\\
%  \mb \beta_{j,1}  &=& \mb \beta_{j,1}, \mbox{for }j = 0,\ldots,p,
\end{eqnarray} 
where $K = C-1$.  Note that $\delta_{j,k}$ is the coefficient associated with feature $j$ ($j=0$ representing intercept) and $y=k+1$.  With $\delta_{j,k}$'s, the model \eqref{eqn:ygivenx} is now written as:
\begin{eqnarray}
P(y_i = k + 1| \mb x_{i,1:p}, \mb\delta_{0:p,1:K}) &=&
\dfrac{I(k = 0) + I(k > 0)\exp\left(\delta_{0k} + \mb x_{i,1:p}
\mb\delta_{1:p,k}\right)}{1 + \sum_{k=1}^K\exp\left(\delta_{0k} + \mb x_{i,1:p}
\mb\delta_{1:p,k}\right)},
\label{eqn:distygivenx-new}
\end{eqnarray}
for $k=0,\ldots,K$, and $i = 1,\ldots,n$.

We can transfer the symmetric prior for $\mb\beta_{j,1:C}$ to $\mb\delta_{j,1:K}$ rather than assigning independent priors for $\mb\delta_{j,1:K}$. Applying the standard transformation results for multivariate normal, the transformed parameters $\mb\delta_{j,1:K}$ are distributed with a joint multivariate Gaussian distribution:
\begin{eqnarray}
  \mb \delta_{j, 1:K} | \sigma_j^2 \sim N_{K}(\mb 0, (I_K + J_K)\sigma_j^2), \mbox{for }j = 0,\ldots,p,
\label{eqn:priordeltas}
\end{eqnarray}
where $I_K$ is a $K\!\times\! K$ identity matrix and $J_K$ is a $K\!\times\! K$
matrix with all elements 1. More explicitly, the joint PDF for 
\eqref{eqn:priordeltas} is given as follows:
\begin{eqnarray}
  P(\mb\delta_{j,1:K}|\sigma_j^2) &=& (2\pi\sigma_j^2)^{-K/2} \exp \left(
  - \dfrac{V(\mb \delta_{j,1:K})}{2\sigma_j^2}\right)\times |I_K + J_K|^{-1/2},
\mbox{where,}\\
V(\mb \delta_{j,1:K}) &=& \sum_{k=1}^K\delta_{jk}^2 -
\left(\sum_{k=1}^K\delta_{jk}\right)^2\Big/C. \label{eqn:vvalue}
\end{eqnarray}
Note that for $C=2$, \eqref{eqn:priordeltas} is just a
univariate normal for $\delta_{j,1}$ with variance $2\sigma_j^2$. 

We see that $V(\mb \delta_{j,1:K})$ in \eqref{eqn:vvalue} is the sum of squared differences of $(\mb 0,\delta_{j,1},\ldots,\delta_{j,K})$ from its mean $(0+\sum_{k=1}^K\delta_{jk})/C$. $V(\mb \delta_{j,1:K})$ is exactly the same as the sum of squared differences of $\mb\beta_{j,1:C}$ to its mean, that is,  
\begin{equation}
V(\mb\delta_{j,1:K}) =  \sum_{c=1}^{C}(\beta_{j,c}-\bar \beta_{j})^{2}, \mbox{where }\bar\beta_{j} = (1/C)\sum_{c=1}^{C}\beta_{j,c}, \mbox{ for }j = 0,\ldots,p.
\end{equation}
For feature selection, it is more straightforward to look at the standard deviation of $\mb\beta_{j,1:C}$, which is defined as a function of $\mb\delta_{j, 1:K}$:
\begin{equation}
\mbox{SDB}(\mb\delta_{j,1:K})=\sqrt{V(\mb\delta_{j,1:K})/C} , \mbox{for }j = 0,\ldots,p \label{eqn:sdb}
\end{equation}
Note that when $C=2$,  SDB$(\mb\delta_{j,1})=|\delta_{j,1}/2|$.

\subsection{Horseshoe and NEG Priors}
As alternatives to the Inverse-Gamma distribution in \eqref{eqn:priorsgm}, other priors for $\sigma^2_{j}$ have been proposed in the recent literature for regression problems with the goal of shrinking $\mb\sigma_{j}^2$ associated with weak signal more towards 0. For example, \citet{carvalho2010horseshoe} proposed a horseshoe prior for coefficients by assigning a half (positive) Cauchy distribution for $\mb\sigma_{1:p}$. For uniformity of notation, we describe the half-Cauchy distribution using a half-\textit{t} distribution with various degrees of freedom for $\sigma_j$, inducing the following prior for $\sigma_j^2$:
\begin{equation}
 P_{ghs} (\sigma_j^2) = \dfrac{\Gamma ((\alpha+1)/2)}{\Gamma
(\alpha/2)\sqrt{\alpha \pi} \sqrt{w}} \left(\dfrac{1}{1+ \sigma_j^2/(\alpha
w)}\right)^{\frac{\alpha+1}{2}}\, \dfrac{1}{(\sigma_j^2)^{1/2}}, \ \ \mbox{for } \sigma_j^2 > 0.
\label{eqn:priorsgm-hs}
\end{equation}
\citet{griffin2011bayesian} proposes using an NEG prior for coefficients by assigning an exp-gamma prior for $\sigma^2_{1:p}$: $\sigma_j^2 |\psi_j \sim \exp(\frac{1}{\psi_j}), \psi_j \sim \mbox{IG}(\alpha/2, \alpha w/2)$, where $\psi_j$ is the mean parameter of exp distribution. We can marginalize $\psi_j$ and obtain a closed-form PDF for $\sigma_j^2$:
\begin{equation}
P_{neg}(\sigma_j^2) =
\dfrac{\kappa}{\lambda}\left(\dfrac{1}{1+\sigma_j^2/\lambda}\right)^{\alpha/2 +
1}, \ \ \mbox{for } \sigma_j^2 > 0 \label{eqn:priorsgmneg}
\end{equation}
where $\kappa=\alpha/2$ and $\lambda = \alpha w/2$ for notational simplicity. We will call \eqref{eqn:priorsgm-hs} and \eqref{eqn:priorsgmneg} the \textbf{GHS} and \textbf{NEG} priors for $\sigma_j^2$. Note that these names are also used for the priors of the coefficients $\mb\delta_{j, 1:K}$ when $\sigma_{j}$ integrated out.  The PDFs of GHS and NEG priors for $\sigma_j^2$ do not converge to $0$ as $\sigma_j^2$ goes to $0$ (as the IG prior does); therefore it is possible that regression coefficients are better shrunken towards $0$ without punishing large signals. This property is indeed desired, and may be beneficial. However, our numerical experiments show that using these two priors over a $t$ prior makes little difference in MCMC inference. Additionally, adaptive rejection sampling \citep[][]{gilks1992adaptive} (or other sampling methods) for the posteriors of $\sigma_{j}^2$ given $\mb\beta_{j,1:C}$ based on GHS and NEG priors is needed in Gibbs sampling. By contrast, direct sampling for $\sigma_{j}$ is available if IG is used.  This additional sampling could significantly increase the total MCMC sampling time when $p$ is large. 

Choosing the scale $\sqrt{w}$ is an important issue for any inference with shrinkage.  However, when $\alpha=1$, we recommend to fix $w$ around a reasonable value, \textit{e.g.} $\log(w) = -10$.  Most random numbers generated by $t$/GHS/NEG distributions with this setting are very small, but contain a fairly large portion of values between 0 and 2, which can model a wide range of problems.  Our following empirical results will show that when $\alpha = 1$, the fitting results are not sensitive to the choice of  small $w$.   When $\alpha > 1$,  it is better to treat the scale as a hyperparameter assigned with a prior; this is because the results are sensitive to $\sqrt{w}$.  In our implementation, we assign a vague normal prior for $\log (w)$.  
\subsection{Gibbs Sampling Procedure}\label{sec:gibbs}

\iffalse Here we use the term of ``Gibbs sampling'' to mean a general sampling procedure that alternatively ``samples'' conditional distributions of a subset of variables of a full (joint) distribution given the others are fixed. We put quote around the word \textit{sample} because we don't have to \textit{directly} sample from each conditional distribution. Alternatively it can be just a transformation (deterministic or random) of the subset of variables that leaves the conditional distribution invariant (if current state has the distribution, the transformed state still has the same distribution). This transformation needs not be reversible either, nor ergodic. A super-transformation that are composed of a series of such transformations is valid because it also leaves the full distribution invariant. A formal explanation of such Gibbs sampling procedure can be found from \cite{neal:1993} (pages 44 - 46).
\fi

We use a Gibbs sampling procedure to sample the full posterior distribution of BLRHL model. The full posterior distribution is written as: 
\begin{equation}
  P(\mb\delta_{0:p,1:K}, \mb\sigma_{1:p}^2 | \mb D) \propto L(\mb
\delta_{0:p, 1:K})\times P(\mb \delta_{0:p, 1:K} | \mb\sigma_{0:p}^2) \times
P\left(\mb\sigma_{1:p}^2 \,|\, \alpha/2, \alpha\,w/2\right),
\label{eqn:fullpost}
  \end{equation}
where $\mb D$ represents the data $y_i, \mb x_{i, 1:p}$ for $i=1,\ldots, p$ and other fixed values in BLRHL models --- $\alpha, \sigma_0^2$; $L$ is the likelihood function:
$
  L(\mb \delta_{0:p, 1:K}) = \prod_{i=1}^n P(y_i|\mb x_{i, 1:p}, \mb\delta_{0:p,
1:K});
$
the last two parts are the PDFs of priors specified  by \eqref{eqn:priordeltas}, and one of the priors given in \eqref{eqn:priorsgm}, \eqref{eqn:priorsgm-hs}, or \eqref{eqn:priorsgmneg}. We sample the full posterior in \eqref{eqn:fullpost} by sampling the conditional distribution of $\mb\sigma_{1:p}^2$ and $\mb\delta_{0:p,1:K}$ given each other alternately for a number of iterations. If IG prior \eqref{eqn:priorsgm} is used, the Gibbs sampling procedure involves alternating the following two steps:
\newcounter{Lcount}
\begin{list}{\textbf{Step \arabic{Lcount}:}}
{\usecounter{Lcount}\itemindent=0pt\topsep=-5pt}
\sf

\item Given $\mb\sigma_{1:p}^2$ fixed, update $\mb\delta_{0:p, 1:K}$ jointly
with a Hamiltonian Monte Carlo transformation that leaves invariant the
following distribution:
   \begin{equation}
   P(\mb\delta_{0:p,1:K} |\mb\sigma_{0:p}^2, \mb D) \propto 
   L(\mb\delta_{0:p,1:K})\times P(\mb \delta_{0:p, 1:K} | \mb\sigma_{0:p}^2). 
   \label{eqn:postdelta}
   \end{equation}  
\item Given value of $\mb\delta_{1:p,1:K}$ from Step 1, update $\mb
\sigma_{1:p}^2$ by sampling from
   \begin{eqnarray}
    \sigma_{j}^2|\mb\delta_{j,1:K} &\sim& 
   \mbox{IG} \left(\sigma_{j}^2 \,\Big|\,\frac{\alpha+K}{2},
   \frac{\alpha w + V(\mb\delta_{j,1:K})}{2}\right), \mbox{for }j = 1,\ldots,p\label{eqn:postsgm}
   \end{eqnarray}
\end{list}
 
Note that in {\sf \textbf{Step 2}}, $\sigma^{2}_{0}$ is opted out of the updating process as it is fixed at a large value. The sampling for \eqref{eqn:postsgm} in {\sf \textbf{Step 2}} is straightforward; when Horseshoe and NEG priors are used, the sampling method for {\sf \textbf{Step 1}} can be the same, but we have to use the posterior of $\sigma_j^2$ given $\mb\delta_{j,1:K}$ differently in {\sf \textbf{Step 2}}. When we use GHS prior \eqref{eqn:priorsgm-hs} for $\sigma_j^2$, the conditional posterior of $\sigma^2_j$ given
$\mb\delta_{j:1:K}$ [instead of Equation \eqref{eqn:postsgm}] is: 
\begin{equation}
P_{ghs}(\sigma_j^2|\mb\delta_{j,1:K}) \propto
\dfrac{1}{(\sigma_j^2)^{K/2}}\,\exp\left(-\dfrac{\Vj}{2\sigma_j^2}\right)\times
\left(\dfrac{1}{1+ \sigma_j^2/(\alpha w)}\right)^{\frac{\alpha+1}{2}}\,
\dfrac{1}{(\sigma_j^2)^{1/2}}.\label{eqn:postsgm-ghs}
\end{equation}
The induced conditional distribution for  $\xi_j=\log (\sigma^2_j)$ from the above distribution is log-concave and can be sampled with ARS \citep{gilks1992adaptive}. When we use an NEG prior \eqref{eqn:priorsgmneg} for $\sigma_j^2$, the conditional posterior of $\sigma^2_j$ given $\mb\delta_{j,1:K}$ [instead of Equation \eqref{eqn:postsgm}] is
\begin{equation}
P_{neg}(\sigma_j^2|\mb\delta_{j,1:K}) 
\propto
\dfrac{1}{(\sigma_j^2)^{K/2}}\,\exp\left(-\dfrac{\Vj}{2\sigma_j^2}\right)\times
\left(\dfrac{1}{1+\sigma_j^2/\lambda}\right)^{\alpha/2 + 1}. \label{eqn:postsgm-neg}
\end{equation} 
The induced posterior of the log transformation $\xi_j=\log (\sigma_j^2)$ from the above distribution is log-concave and can be sampled with ARS.

The key component in the above procedure is the use of Hamiltonian Monte Carlo (HMC) for updating high-dimensional $\mb\delta_{0:p,1:K}$. In Section \ref{sec:hmc} we give a concise description of HMC; this method can greatly suppress the random walk  due to correlation (which is common in logistic regression posteriors; see our real data examples) with a long leapfrog trajectory \citep{neal2010mcmc}. The major problem of sampling from  posteriors based on hyper-LASSO priors is the existence of many local modes due to feature redundancy. Applying HMC in the above Gibbs sampling framework can travel across the modes fairly well for the following reason. When both $\sigma^2_j$ for two correlated features are large, the joint conditional distribution of their coefficients given $\sigma_j^2$ in {\sf \textbf{Step 1}} is highly correlated, probably close to their likelihood function as shown in Figure \ref{fig:killcor}. Because a fairly long HMC trajectory has a much greater chance than ordinary MCMC methods to move from one end of the contour to the other end, the Markov chain can travel from one mode to another.  For high-dimensional problems with very large $p$, such as thousands, {\sf \textbf{Step 1}} is computationally intensive as it involves updating $p*K$ coefficients in each step. This challenge can be relieved greatly by an important trick that we call ``restricted Gibbs sampling'', where only the coefficients with $\sigma_j^2$ greater than a certain threshold are updated in {\sf \textbf{Step 1}}. The details of this trick are given in Section \ref{sec:appendix}.  A list of notations for the settings of BLRHL is given in Section \ref{sec:settings}.

%Very long Markov chains (1M iterations of Gibbs sampling with HMC trajectory of length 50) for applications to prostate microarray data with 6033 genes took about 10 hours. Such amount of time may be tolerable for most scientific research problems. It is also worth mentioning that, for getting better results using our MCMC methods, one needs only to run Markov chain longer without using other mysterious tricks. MCMC method is indeed more convenient, though slow. 

\subsection{Feature Importance Measures from Markov chain Samples} \label{sec:ranking}

With posterior samples of $\mb\delta_{1:p, 1:K}$, we recommend using \textit{means} over iterations to estimate the coefficients (these estimates are denoted by $\hat{\delta}_{j,1:K}$). We then compute SDB($\hat{\mb\delta}_{j,1:K})$ using formula \eqref{eqn:sdb} to obtain an importance measure for feature $j$; these features can then be ranked by SDB($\hat{\mb\delta}_{j,1:K})$. As we have discussed in Section \ref{sec:simple}, the Markov chain sample pool is a mixture of subpools from different modes, each corresponding to a succinct feature subset.  Thus, the mean over the Markov chain is  a summary of the importance of the feature, not an estimate of the true coefficient.  However, this method omits some useful features that appear with low frequency in Markov chain samples. In the context of high-dimensional problems, there are often a large number of such correlated features, therefore, discriminating them according to their predictive ability is desired. The ranking by \textit{means} can omit many correlated features with weaker predictive ability as well as those totally useless, hence pinning down a very small subset of highly relevant features. 
%In light of the multi-modality in posterior, the means may not be the best choice for interpreting the MCMC simulation results, especially in the datasets with a large number of highly correlated features.  However, in this paper we wish to focus on discussing the choice of heaviness and scale of hyper-LASSO priors. We will discuss other alternatives in the Section \ref{sec:end}.  

\section{Simulation Studies} \label{sec:sim} 
 
\subsection{Comparing Scaling Effects in LASSO and Hyper-LASSO} \label{sec:sim100}

\begin{figure}[ht]
 \centering
  \caption{Scatterplots of the first three features from a simulated dataset used in Section \ref{sec:sim100}.}
 \includegraphics[height=0.25\textheight, width=0.9\textwidth]{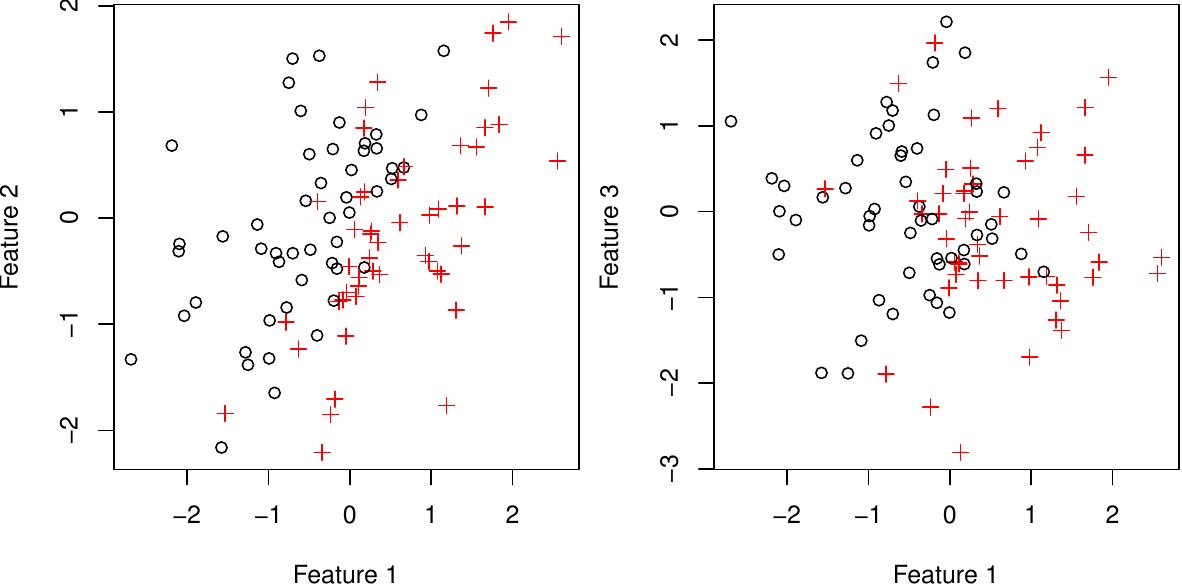}
 \vspace*{-10pt}
 \label{fig:datap100}
\end{figure}

We  generated a dataset of $n=1100$ cases (of which 100 were used for fitting models and the other 1000 were used to look at predictive performance) and $p = 200$ features, where the response $y_{i}$ is equally likely to be 1 and 2. Given $y_{i}$,  200 features are generated from the following Gaussian models:
  \begin{eqnarray}
x_1|y=c &=& \mu_{c} + z_1 + \epsilon_1, \\
x_2|y=c &=& 2z_1 + z_2 +\epsilon_2,\\
x_{j}|y=c &=&\epsilon_{j}, \mbox{  for  } j = 3, \ldots, 200
\label{eqn:sim1x}
  \end{eqnarray}
where $\mu_{1} = 0, \mu_{2}=2$ and $z_1, z_2,\epsilon_j$ are all distributed as $N(0,1)$. In this model, only the first feature is differentiated across two classes, and the 2nd  is non-differentiated but correlated with the 1st; therefore only the first two features are useful for predicting the response $y$. Using Bayes rule, we can find the conditional distribution of $y$ given $\mb x$; this distribution is a logistic regression model with \textbf{true coefficients} $\mb \delta_{0:2,1} = (0, 2.60, -1.22)$ and others equal to $0$. The relationship between $y$ and $\mb x$ is simple, but the signals are placed among the other 198 unrelated features. Figure \ref{fig:datap100} shows the scatterplots of the 2nd and 3rd features to the 1st with shapes representing the two classes. 

\begin{figure}[ht]
 \centering

\caption{Comparison of coefficient estimates and AMLPs against log scale for the study in Section \ref{sec:sim100}. The numbers on the top  of \ref{fig:p200pathes} show the number of coefficients with absolute values not smaller than 0.1 times the maximum value in all coefficients.  }

  \subfloat[][Coefficient paths of BLRHL (left) and LASSO (right).  \label{fig:p200pathes}]
{
 \includegraphics[trim=0 0 0 20pt, clip,width=0.95\textwidth,height=0.3\textheight]{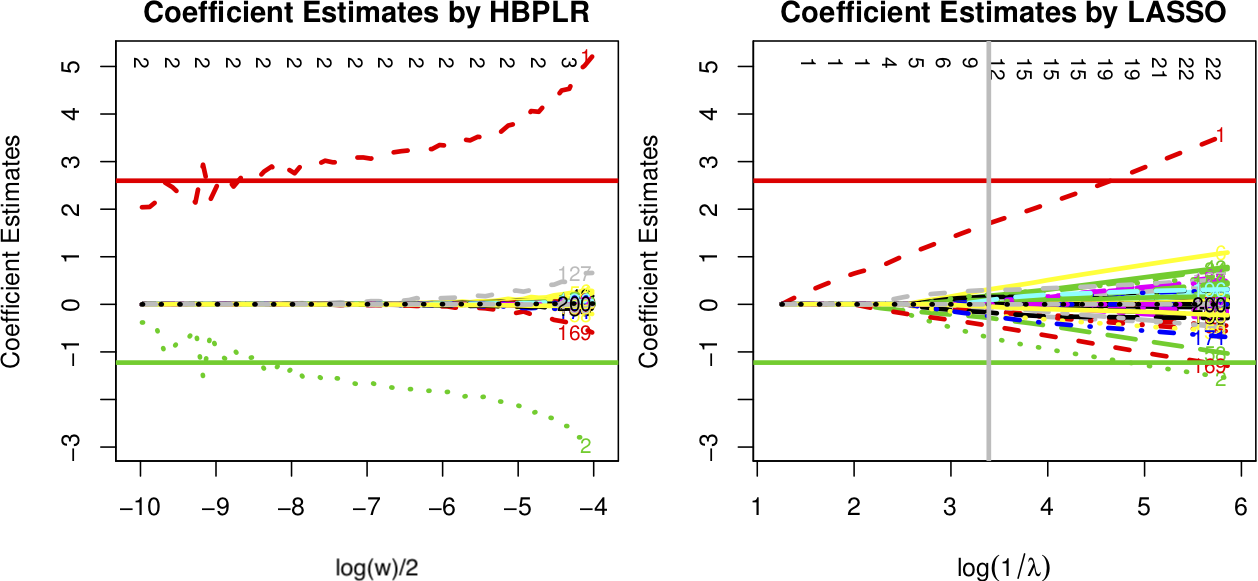}
 
}

  \subfloat[][AMLPs of BLRHL (left) and LASSO (right). 		
  \label{fig:p200predpath}]{
  \includegraphics[trim=-30pt 0 0 14pt, clip,width=0.95\textwidth,height=0.3\textheight]{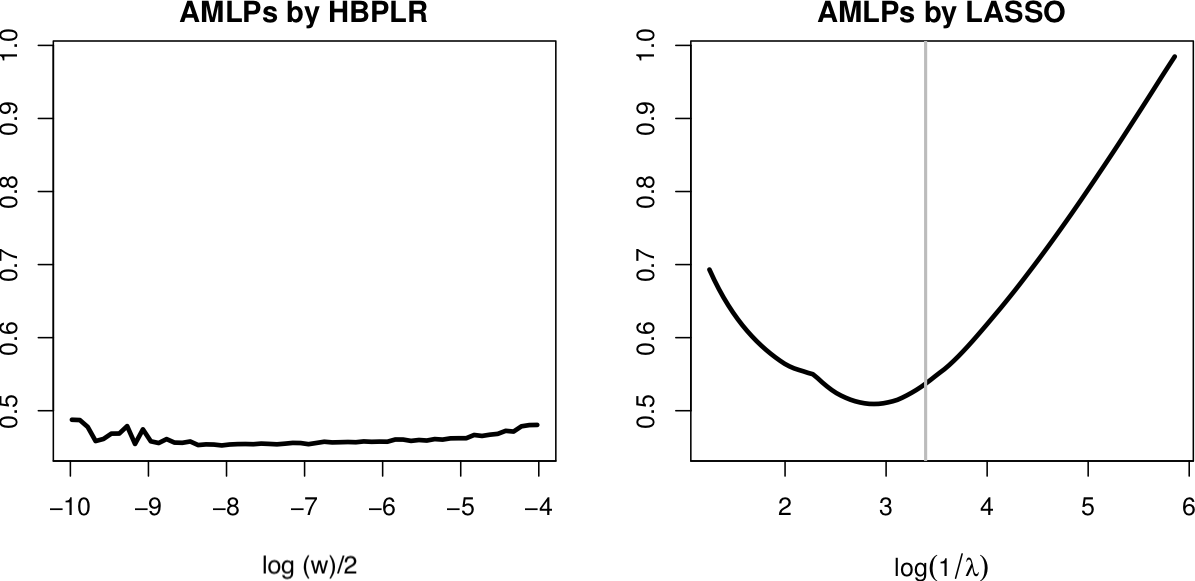}
  }
%  
%  \subfloat[][Error Rate]{
%  \includegraphics[trim=0 0 0 14pt, clip,width=0.95\textwidth,height=0.23\textheight]{images/blers.png}
%  }
% 
\vspace*{-20pt}
\end{figure}

We ran BLRHL that uses a \textit{t} prior with $\alpha = 1$ and MCMC settings as follows: $n_1=50K, \ell_1 = 5, n_2=100K,  l_2=50, \epsilon = 0.3, \zeta = 0$, and 100 different $\log(w)$ spaced evenly from $-24$ to $-8$. \textit{The meanings of these setting parameters are listed in Section \ref{sec:settings}}. %The Markov chains are unnecessarily long as the dimension is very small, but we used it since the computation is fast, taking about 30 mins for each chain. 
For each choice of scale, we estimated the coefficients using \textit{means} of Markov chain samples. These results allow us to draw the solution paths (Figure \ref{fig:p200pathes}) of all the coefficients against the log of the scale and compare the path given by LASSO (using R package \texttt{glmnet}).  We can see that BLRHL gives much more distinctive estimates  of the two non-zero coefficients from those of the other 198 useless features than LASSO. Due to the inclusion of many useless features, LASSO cannot identify the second feature distinctively. From comparing these paths, we also see that the estimates by BLRHL are very stable for the choices of $w$ in a very wide range. There is an upward bias in the mean estimates when $\sqrt{w}$ is large; this is because the marginal posterior distributions of the two coefficients for two correlated features are skewed to large absolute values. This bias, however, does not affect the predictive performance and feature selection. 

Figure \ref{fig:p200predpath} shows the predictive performance of BLRHL and LASSO measured by AMLP --- \textit{the average minus log predictive probabilities at the true labels}.  The AMLP paths are shown in Figure \ref{fig:p200predpath}. We see that BLRHL predicts better than LASSO; more importantly, the predictive performance of BLRHL is very stable for a wide range of $\sqrt{w}$. By contrast, LASSO is very sensitive to the choice of $\sqrt{w}$. %if a small scale is used, LASSO omits more useless features, but also over-shrinks the really large coefficients, especially weaker $x_2$, therefore gives poor prediction; conversely, if a large scale is used, LASSO takes many ``hay'' into the model, which also downgrades the predictive performance. We see that the cross-validated choice based on the small training dataset did not pick up the optimal $\log(1/\lambda)$, around 3; and even when this optimal choice is made, LASSO nearly omits $x_2$. BLRHL does not have this difficulty, and can identify the 2 ``needles'' from the 198 ``hay''.

%\begin{figure}[ht]
% \centering
%  \caption{Coefficient paths of BLRHL (left) and LASSO (right). The numbers on the top show the number of coefficients with absolute values not smaller than 0.1 times the maximum value in all coefficients. The vertical line in LASSO path shows the cross-validation choice of $\log(1/\lambda)$. The numbers beside paths on the right are feature indice.}
%  \vspace*{-5pt}
% 
%\vspace*{-15pt}
%\end{figure}
%
% 

\subsection{Investigating the Choice of Heaviness ($\alpha$)}\label{sec:sim2000}

We generated 50 datasets ($n=2100$ cases, $100$ of which was used for training, while the other 2000 were used to test predictions) as follows. The number of classes $C$ is set to 3, and class labels are equally likely drawn from 1,\,2, and 3. The first two features were generated in a similar way as the $x_{1}$ and $x_{2}$ of the dataset used in Section \ref{sec:sim100}, with an addition of class $3$ having $0$ for means.  We add another group of features ($x_{3} - x_{10}$) that are highly correlated within groups but are independent of $x_{1}$ and $x_{2}$, and have mean equal to 2 in class 3.  More specifically,  values of these 10 features for each case were generated as follows:\\
\hspace*{-20pt}\begin{minipage}{4.3in}
\begin{eqnarray*}
x_1|y=c &=& \mu_{c,1} + z_1 + 0.5 \epsilon_1,\\
x_2|y=c &=& \mu_{c,2} + 2z_1 + z_2 + 0.5 \epsilon_2,\\
x_j|y=c &=& \mu_{c,j} + z_3 + 0.5 \epsilon_j, \ \mbox{for }j = 3,\ldots,10,
\end{eqnarray*} 
\end{minipage}

\vspace*{-1.1in}\hspace*{2.9in}\begin{minipage}{4.25in}
\begin{equation*}
\mbox{where }(\mu_{c,j})_{3\times 10} = 
\left(
\begin{array}{ccccc}
0 & 0 & 0 &\ldots &0\\
2 & 0 & 0 & \ldots &0\\
0 & 0 & 2 & \ldots &2
\end{array}
\right),
\end{equation*}
\end{minipage}

\bigskip

\noindent and $z_j$ and $\epsilon_j$ are independently generated from $N(0,1)$. In addition to these 10 features, we also attached 1990 features simply drawn from N(0,1), which we will call absolute ``hay''.  In this model, $x_1$ is differentiated with a different mean in class 2 from classes 1 and 3. $x_2$ is non-differentiated, but correlated with $x_1$ and therefore is useful, as shown by Figure \ref{fig:datap100}. $x_3-x_{10}$ are all differentiated, with different means in class 3 from classes 1 and 2. However $x_3-x_{10}$, which have the same class means and are related to a common factor $z_3$, are highly correlated and redundant for predicting $y$; we will refer to this group as ``correlated features''. 

%Here we use $p=2000$ for experimental time consideration.  Our methods can be applied to problems of higher dimensions, such as more than 6000 features in the case study presented in Section \ref{sec:gene}. However, generally, we think that in practice, we should apply BLRHL to a much smaller subsets of features pre-selected with univariate screening to discard those obviously useless features (such as those with very small variation at all), rather than on the original dataset with millions of features.  

% We consider that identifying at least one from the correlated features and also discriminating them according to their predictive ability is a good property for high-dimensional feature ranking techniques.

We ran BLRHL using \textit{t} priors with 4 choices of $\alpha$: 0.2, 0.5, 1, 4, 10; the  setting for $\log (w)$ varies slightly for different $\alpha$. When $\alpha=1$, we chose two values of $\log(w)$: -20 and -10.  When $\alpha=4$ and 10, we chose to treat $\log(w)$ as a hyperparameter assigned with a normal prior with variance 100 (this is because for large $\alpha$, the results for feature selection and prediction are sensitive to the choices of scale and we therefore show the results with $\log(w)$ chosen automatically during MCMC simulation). The values of $\log(w)$ for $\alpha = 0.2/0.5$ are -40/-20 respectively. For setting MCMC, when $\alpha = 1$, we chose $n_1=50K, \ell_1=10, n_2=500K, \ell_2=50, \epsilon=0.3, \zeta = 0.05$.  The details of these setting parameters can be found in Appendix \ref{sec:settings}. In particular we have run MCMC for various choices of $\zeta$ in restricted Gibbs sampling for a given dataset; the results are fairly stable.  When $\alpha$ equals to 0.2/0.5/4/10 (other than 1), we set a larger $n_2 = 1M$ for the longer chain, with the other settings being the same as the $\alpha=1$ case.  We ran BLRHL using GHS and NEG priors  with settings $\alpha = 1$ and $\log (w) = -10$, and the same other settings for running BLRHL using a \textit{t} prior with $\alpha = 1$. We ran LASSO with $\lambda$ chosen by cross-validated AMLP.

\begin{figure}[p]
    \centering
    \caption{SDBs given by BLRHL methods and LASSO on a synthetic dataset with $p=2000$ features. The red numbers show the indices of top  features with relative SDB greater than 0.1, except for (\ref{fig:bsdb100}) which shows indices of features with relative SDB greater than 0.5 (to avoid showing too many indices). The horizontal lines indicate the values equal to 0.1 and 0.01 times the maximum SDB. The x-axis is in log-scale in order to better look at signal features.}\label{fig:p2ksdbs}
    
    \vspace*{-5pt}
    
     \subfloat[SDBs of LASSO]{
    \includegraphics[trim=0 0 0 23pt, clip,width=0.5\textwidth, height=0.27\textheight]{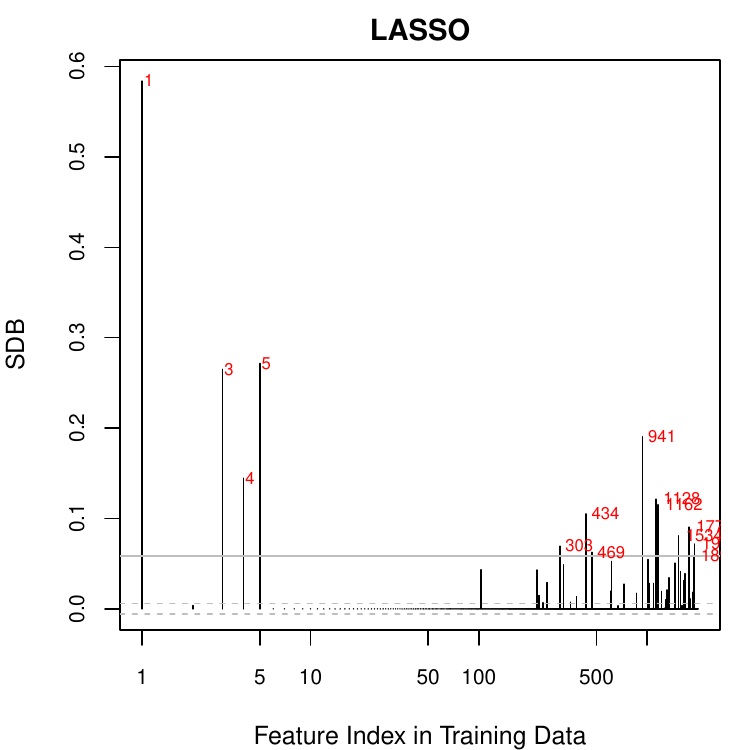}
    }
    \subfloat[SDBs of BLRHL with $t(\df = 10)$]{\label{fig:bsdb100}
    \includegraphics[trim=0 0 0 23pt, clip,width=0.5\textwidth,height=0.27\textheight]
    {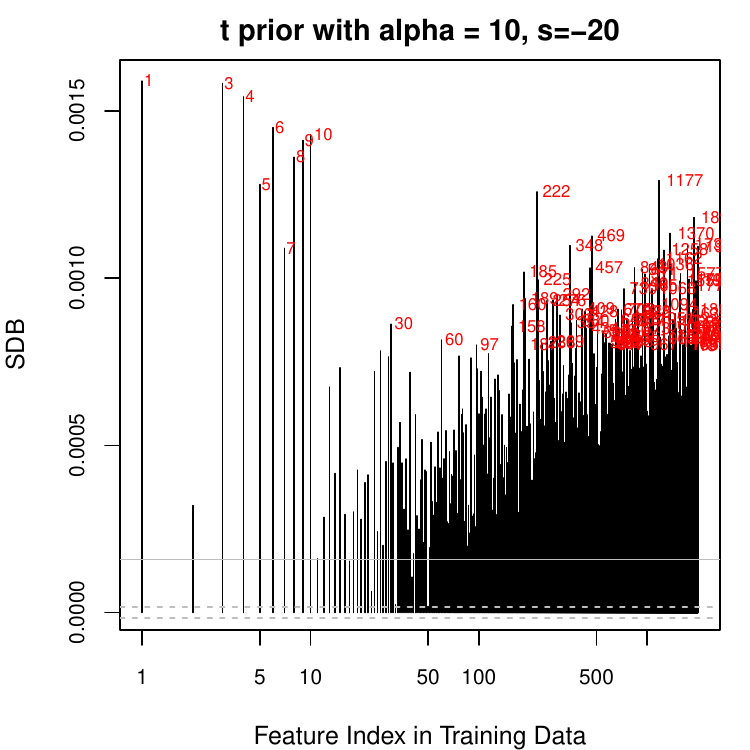}
    }

	\vspace*{-5pt}

    \subfloat[SDBs of BLRHL with $t(\df = 1, \log(w)=-10)$]{ 
    \includegraphics[trim=0 0 0 23pt, clip,width=0.5\textwidth,height=0.27\textheight]{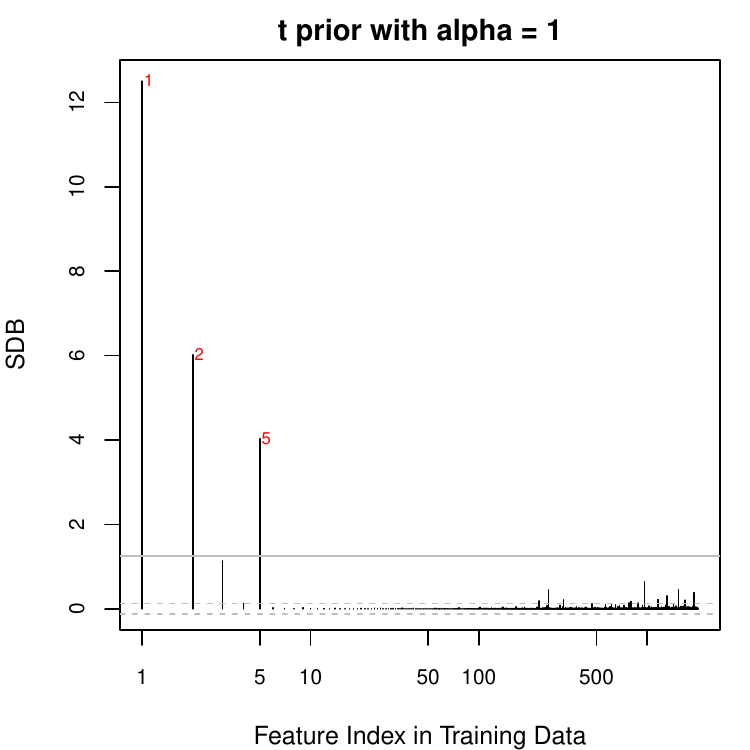} 
    }
    \subfloat[SDBs of BLRHL with $t(\df = 0.2, \log(w)=-40.0)$]{
    \label{fig:bsdb2}
    \includegraphics[trim=0 0 0 23pt, clip,width=0.5\textwidth,height=0.27\textheight]{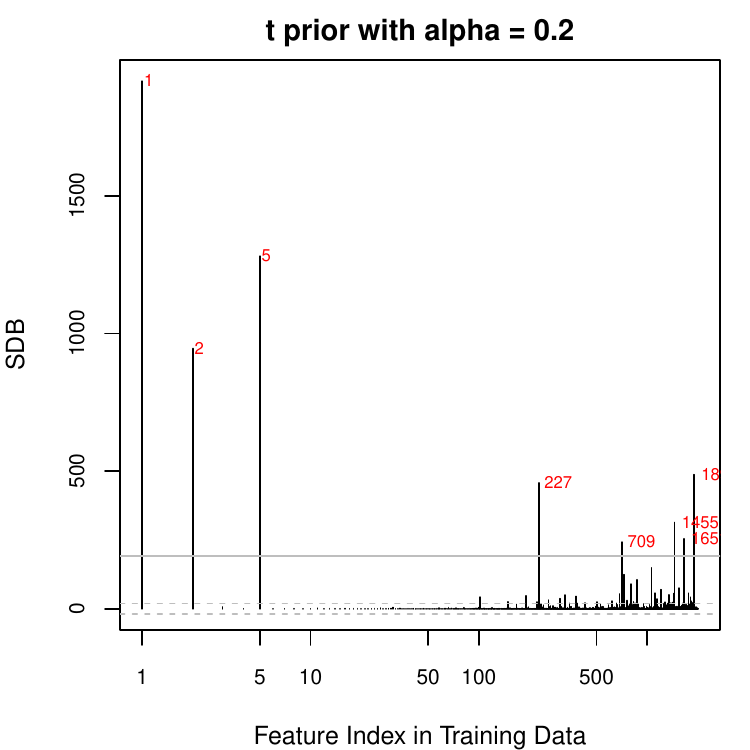}
    }
   
\vspace*{-5pt}

    \subfloat[SDBs of BLRHL with $\mbox{NEG}(\df = 1, \log(w)=-10)$]{ 
    \includegraphics[trim=0 0 0 21pt, clip,width=0.5\textwidth,height=0.27\textheight]{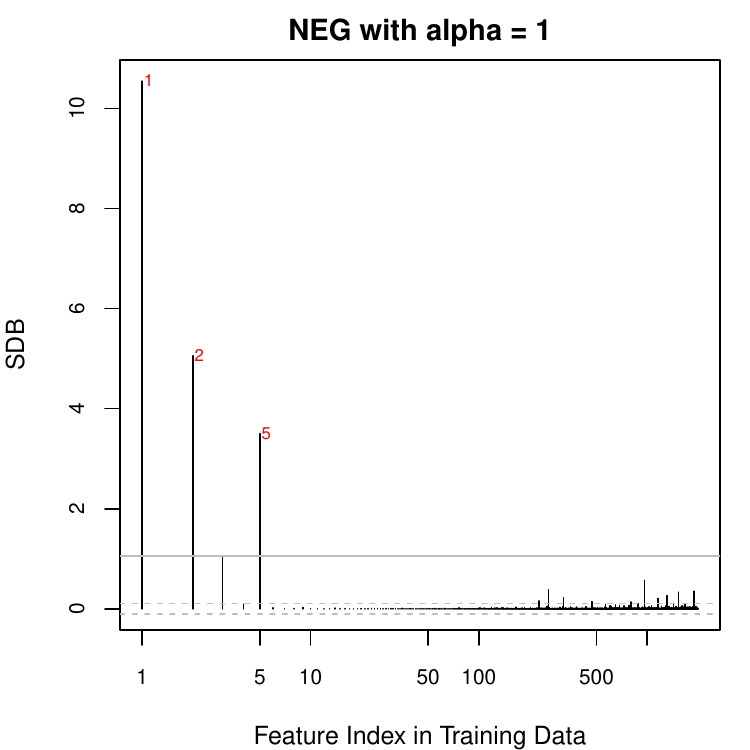} 
    }
    \subfloat[SDBs of BLRHL with $\mbox{GHS}(\df = 1, \log(w)=-10)$]{
    \includegraphics[trim=0 0 0 21pt, clip,width=0.5\textwidth,height=0.27\textheight]{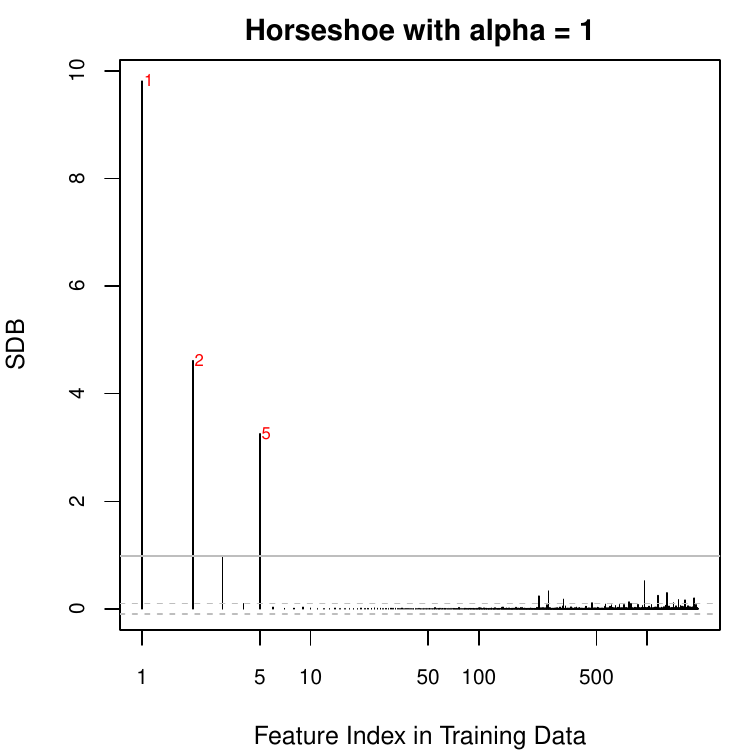}
    }

\end{figure}

We first look at the coefficient shrinkage effects of hyper-LASSO priors with different $\alpha$ and LASSO. Figures \ref{fig:p2ksdbs} show the SDBs of all 2000 features from the dataset.  From Figure \ref{fig:p2ksdbs}, we see that BLRHL methods using \textit{t}/GHS/NEG priors with $\alpha = 1$ perform feature selection very well.  First, they can distinctively separate the absolute ``hay'' from other useful features. Second, they do not miss the 2nd feature which is useful but has weaker relevance. Third, they rank highly one feature ($x_5$) from the 8 correlated features, recognize another feature $x_3$ as useful too, and suppress others. By contrast, when $\alpha$ is large (\textit{e.g.} 10), we see that the LASSO and BLRHL methods: 1) cannot separate the absolute ``hay'' distinctively from the few ``needles''; 2) miss $x_2$ for this dataset (and very often for other datasets), which we think is because they include too much ``hay'' and overfit the data, making $x_2$ harder to identify; 3) tend to include many of the correlated features into their unique mode without clear discrimination for importance. BLRHL with very small $\alpha=0.2$ (very heavy tails) can do feature selection well, but their overly flat tails allow the ``needles'' to go to very large values (such as \textit{thousands}, see Figure \ref{fig:bsdb2}), resulting in very poor prediction in some cases. To summarize the performance of feature selection, we cut SDBs by 0.1 times the maximum SDB (\textit{i.e.}, by thresholding relative SDBs with 0.1).  Table \ref{tab:p2knsel} shows the averages of numbers of retained features in each of the 4 different groups. BLRHL with 10 degrees of freedom selects significantly more noise features; this is because the SDBs of all features are very close due to the light tail (as shown by Figure \ref{fig:bsdb100}). Table \ref{tab:p2knsel} confirms the above observations about the effects of priors with different heaviness in coefficient shrinkage.  The choice of 0.1 as a threshold is an ad-hoc choice; the comparison of feature selection performance between different priors is very similar for different thresholds used to cut the SDBs --- Section \ref{sec:500datasets} presents the feature selection results against a set of choice of thresholds ranging from 0.01 to 0.2 using 500 datasets.

\begin{table}[ht]
\centering
\caption{Means of numbers of retained features by thresholding relative SDBs with 0.1 in different groups in 50 datasets. Numbers in brackets show the standard deviations of the 50 numbers.}\label{tab:p2knsel}
\begin{tabular}{l|lccc}
     & \multicolumn{4}{c}{\textbf{Groups of Features}}\\[5pt]\hline
\textbf{Methods}                           & $x_1$ & $x_2$ & $x_3-x_{10}$ & $x_{11}-x_{2000}$
\\\hline
LASSO                                      & 1     & 0.34 & 2.72 (1.18) & 6.92 (4.97) \\
BLRHL with {\em t}\,(\df=10)               & 0.96  & 0.66 & 7.42 (1.86) & 1354 (580) \\
BLRHL with {\em t}\,(\df=4)                & 1     & 0.36 & 1.26 (0.53) & 0.00 (0.00)\\ 
BLRHL with {\em t}\,(\df=1,$\log(w)=-20$)  & 1     & 0.94 & 1.14 (0.35) & 0.16 (0.37)\\ 
BLRHL with {\em t}\,(\df=1,$\log(w)=-10$)  & 1     & 0.96 & 1.10 (0.30) & 0.32 (0.55)\\ 
BLRHL with GHS\,(\df=1,$\log(w)=-10$)      & 1     & 1.00 & 1.14 (0.35) & 0.30 (0.51)\\ 
BLRHL with NEG\,(\df=1,$\log(w)=-10$)      & 1     & 1.00 & 1.06 (0.24) & 0.28 (0.50)\\
BLRHL with {\em t}\,(\df=0.5,$\log(w)=-20$)& 1     & 0.98 & 1.16 (0.37) & 1.14 (0.97)\\ 
BLRHL with {\em t}\,(\df=0.2,$\log(w)=-40$)& 1     & 0.72 & 1.36 (0.60) & 5.74 (3.12)
\end{tabular}
\vspace*{-5pt}
\end{table}

\begin{figure}[htp]
    \centering
        \caption{Boxplots of AMLPs for 2000 test cases using BLRHL with various priors, and LASSO. ``df'' in the plot represents the choice of $\alpha$ for BLRHL priors. }\vspace*{-5pt}
        \includegraphics[width=0.9\textwidth, height=0.3\textheight
         ]{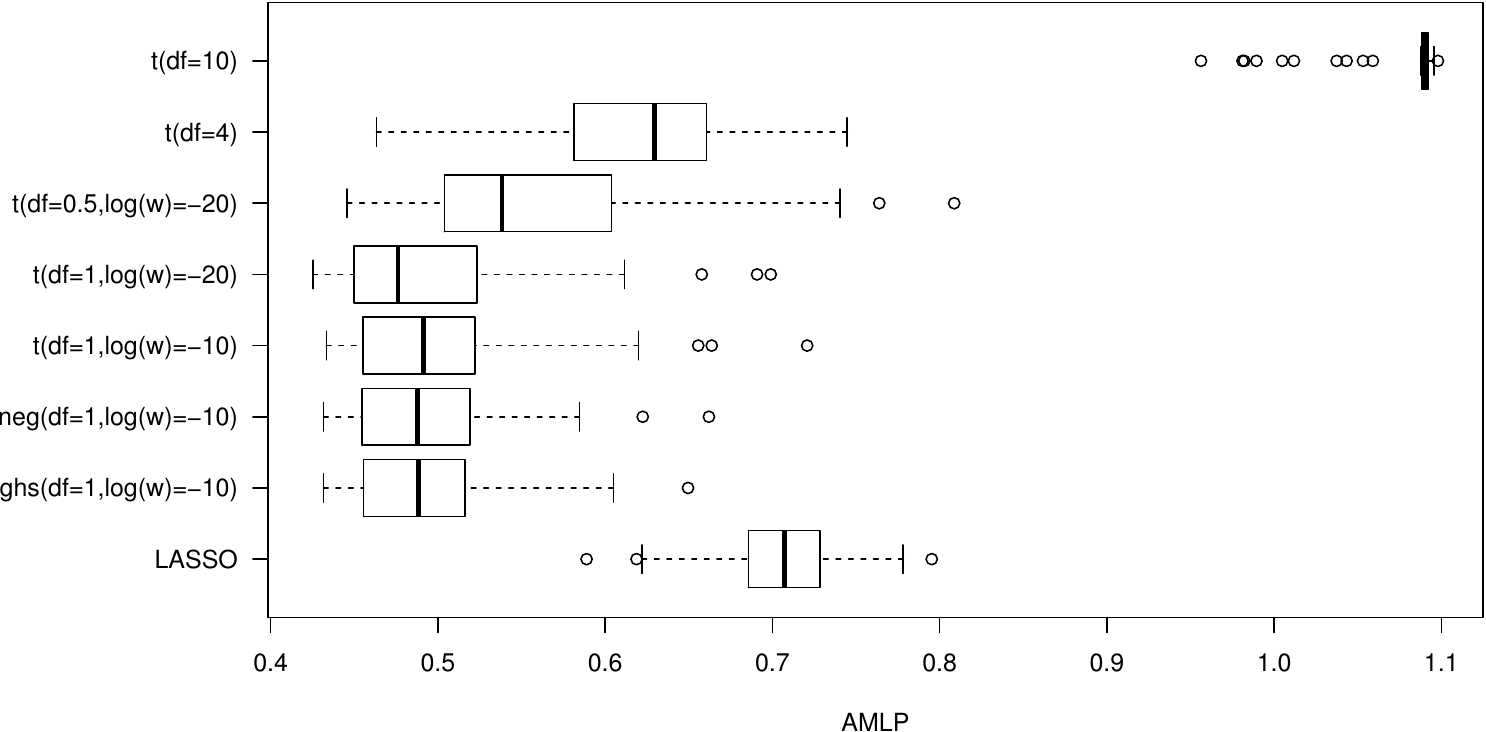}
        \label{fig:p2kamlps}
        \vspace*{-15pt}
        
\end{figure}

Figure \ref{fig:p2kamlps} shows boxplots of the 50 AMLPs for each method on 2000 test cases. From these plots, we see that BLRHL (using {\it t}/GHS/NEG priors) with $\alpha = 1$ gives substantially better predictions for most of the datasets than the other choices of $\alpha$ (as well as LASSO). LASSO and BLRHL with very large and very small $\alpha$ do not predict well. Note that the AMLPs of all runs with $\alpha=0.2$ are \textit{infinity}; these are not shown in Figure \ref{fig:p2kamlps}. Therefore, the choice of $\alpha$ is critical for BLRHL to work well for high-dimensional classification, and $\alpha=1$ is recommended based on our investigation in these simulated datasets with super-sparse signals.

\subsection{Evaluation of BLRHL with 500 Simulated Datasets}\label{sec:500datasets}

We generated 500 datasets (in the same way as described in Section \ref{sec:sim2000}) to evaluate BLRHL more intensively. We apply BLRHL to these datasets using $t$/GHS/NEG priors with $\alpha = 1, \log (w)=-10$ (the optimal choice from the previous investigation), and the same other MCMC settings and compared them to LASSO. 

\begin{figure}[htbp]
\begin{center}

\caption{Comparison of feature selection with 500 datasets}\label{fig:sim500-fdr}
{\includegraphics[width=\textwidth,height=0.55\textheight]{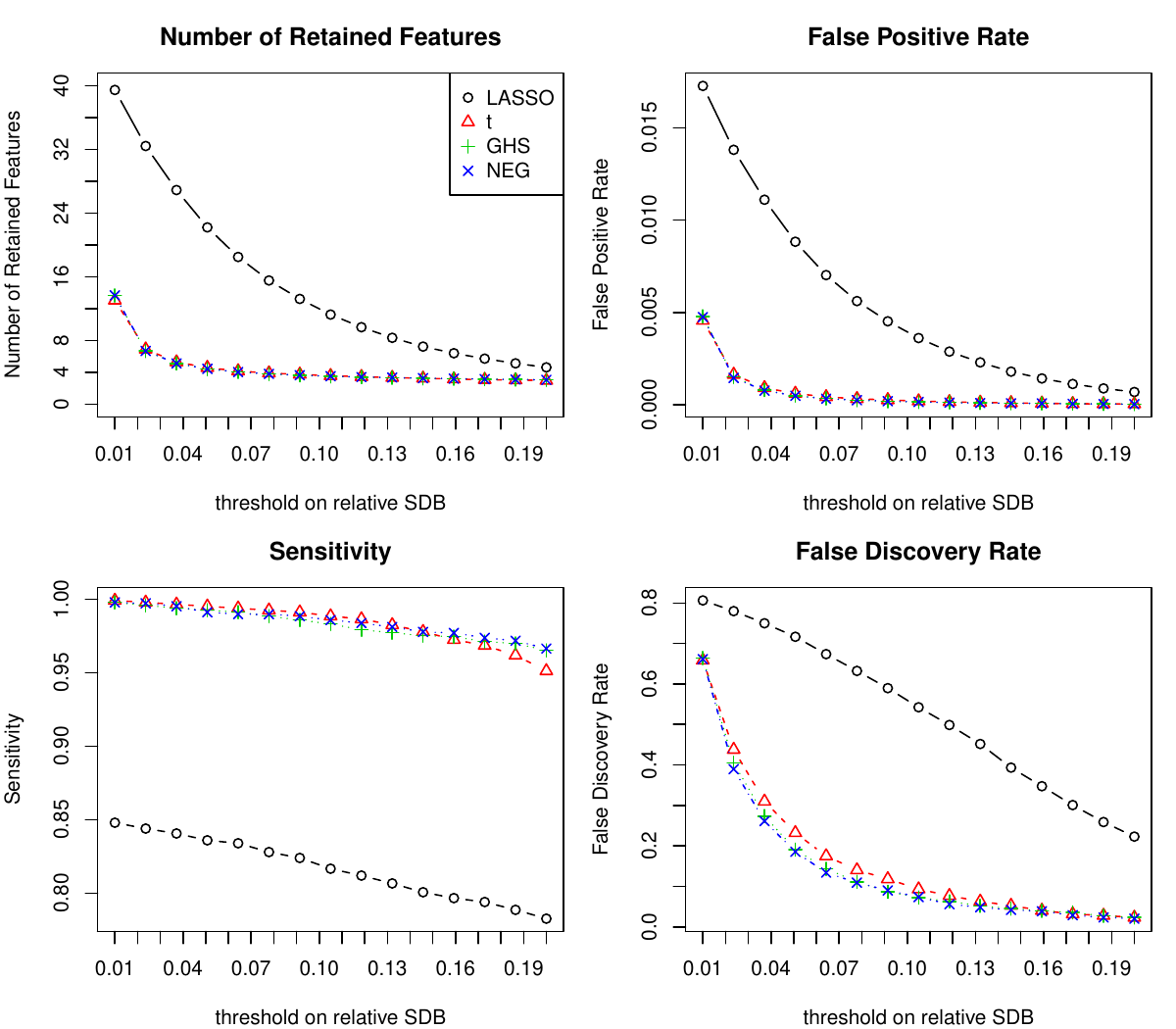}
}
\end{center}
\vspace*{-20pt}
\end{figure}

For each dataset, we perform feature selection by cutting the relative SDBs produced by each method for each dataset using 15 values evenly spaced between 0.01 and 0.2.  At each threshold, we calculate the number of retained features, false positive false rate (FPR), sensitivity (proportion of useful features included),  and false discovery rate (FDR). FDR is defined as the proportion of unrelated features within retained features. In calculating sensitivity,  we treat the features in group 3 (i.e., $x_3$ to $x_{10}$) as a single useful feature. We average the previous four measures over 500 datasets, with results shown in  Figure \ref{fig:sim500-fdr}. Clearly, we see that  BLRHL methods  retain much smaller feature subsets with the same threshold compared to LASSO. We also see that BLRHL has higher sensitivities, lower FPRs, and lower FDRs than LASSO.  The high FPRs and FDRs of LASSO result from the inclusion of many unrelated features. Additionally, Figure \ref{fig:sim500-fdr} indicates that horseshoe and NEG priors have slightly lower FDRs than the $t$ prior. 

\begin{figure}[htbp]
\caption{\footnotesize Comparison of predictive performance with 500 datasets. (a) shows the AMLP boxplots for each method; the top line shows the expected AMLP ($\log(3)$) when one makes random prediction. (b) shows the box plots of percentages of AMLP reductions in comparing BLRHL to the LASSO.}\label{fig:sim500}
\centering

\subfloat[AMLPs\label{fig:sim500-amlp}]
{\includegraphics[width=0.45\textwidth,height=0.3\textheight]{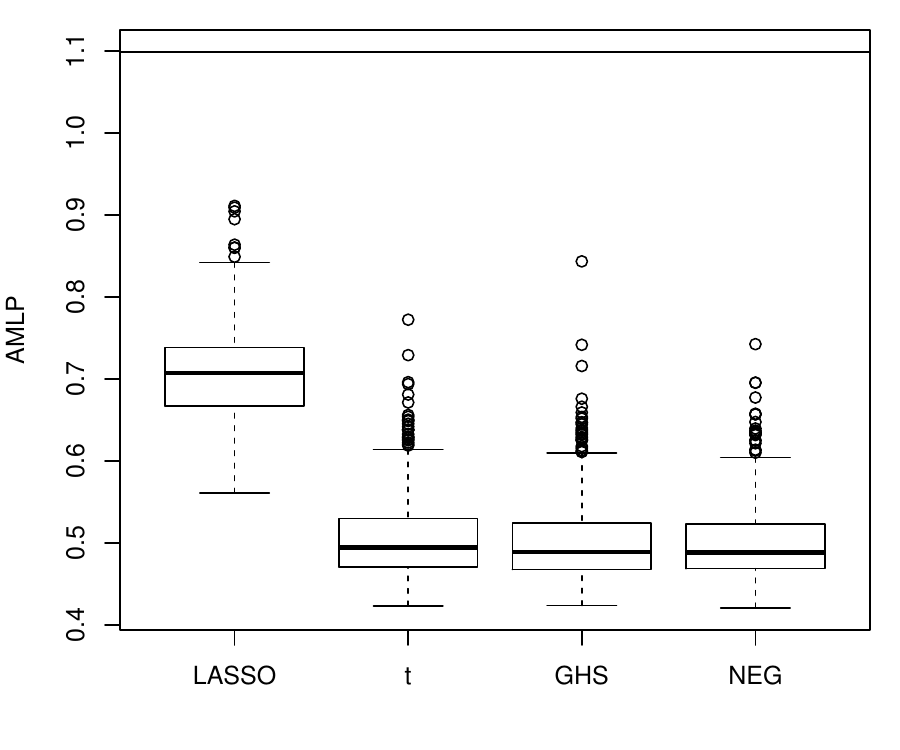}
}
\subfloat[Percentages of AMLP Reduction \label{fig:sim500-amlp-perc}]
{\includegraphics[width=0.45\textwidth,height=0.3\textheight]{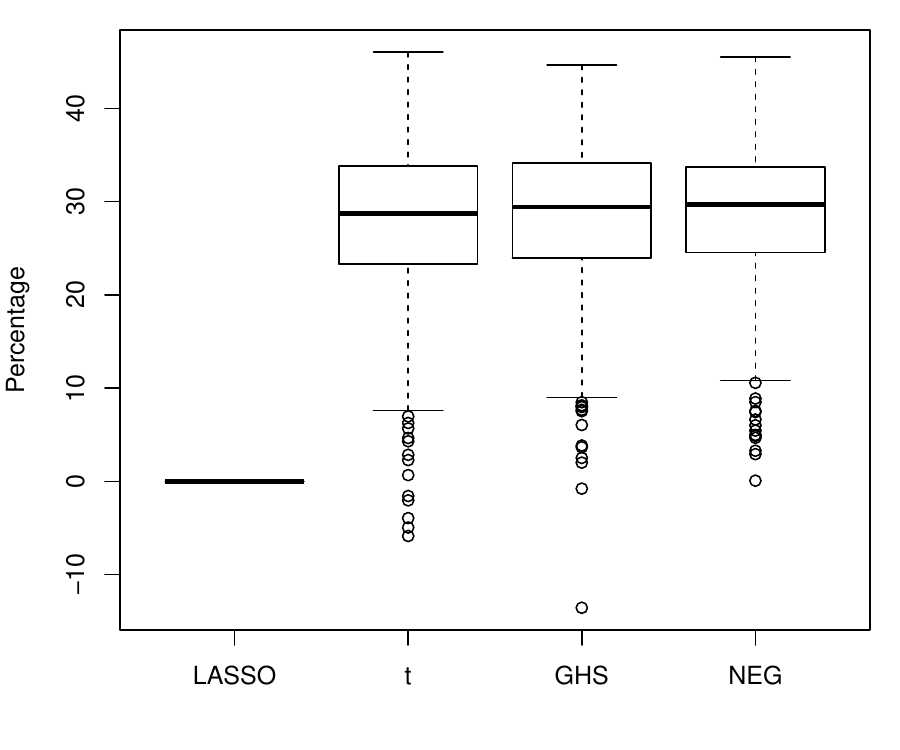}
}

\vspace*{-20pt}
\end{figure}

To compare predictive performance, we collect the AMLPs over 500 datasets for each method. The comparative boxplots of the AMLPs for the four methods is shown in Figure \ref{fig:sim500-amlp}. We see that BLRHL based on different priors with the same $\alpha = 1$ achieves significantly lower AMLPs than LASSO and performs very similarly with each other.  In order to account for the different predictive difficulties in the 500 datasets, we calculated the percentage of AMLP reduction of each BLRHL method relative to the AMLP of LASSO for each dataset (shown in Figure \ref{fig:sim500-amlp-perc}). From this Figure we see that the predictive accuracies of the BLRHL methods are about 30\% higher than LASSO for the majority of these 500 datasets.

\section{Application to a Prostate Microarray Dataset} \label{sec:gene}

We applied BLRHL to a real microarray gene expression data that is related to prostate cancer; this dataset has expression profiles for 6033 genes from  50 normal and 52 cancerous tissues, and was originally reported by \cite{singh2002gene}. We analyzed a dataset downloaded from the website \url{http://stat.ethz.ch/~dettling/bagboost.html} for \citet{Dettling04}, which contains more descriptions about this dataset. To improve the visualization of our results, we re-ordered the features by their F-statistics on the whole dataset; therefore the indices of the genes discussed below are also the ranks of features according to their F-statistics. We ran BLRHL using \textit{t}/GHS/NEG priors and LASSO with $\lambda$ chosen with cross-validation in training cases.  Before fitting with BLRHL, we standardized the features solely with training data (LASSO does such standardization as well). We ran BLRHL with the following settings for each of the 6033 genes: $\alpha = 1, \log(w)= -10, n_1=100K,\ell_1=10, n_2=1M, \ell_2=50, \epsilon = 0.3, \zeta = 0.05$. Each Markov chain took about 10 hours if a \textit{t} prior was used, and about 33 hours if GHS/NEG priors were used. 

%\vspace*{-20pt}

\begin{figure}[t]
\centering
\caption{Gene selection results on prostate data using different methods. For LASSO and BLRHL, the red numbers under points show the indices of ranked genes by their F-statistic, and the horizontal lines indicate the values equal to 0.1 and 0.01 times the maximum SDB. The red numbers show indices of genes selected by thresholding relative SDBs with 0.01. For p-values, the lines indicates -log(0.05) and -log(0.01). The x-axis is in log-scale to better highlight the top genes as selected by the F-statistic rank.}

\subfloat[-log(p-values) given by F-statistic]{ \label{fig:fstat}
  \includegraphics[trim=0 0 10pt 24pt, clip,width=.5\textwidth, height = 0.25\textheight] {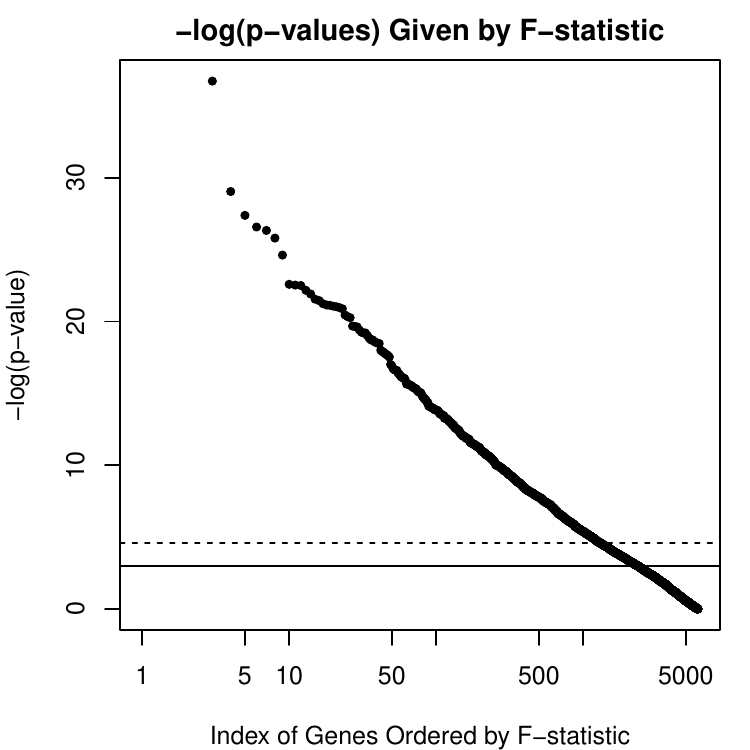}
}
\subfloat[SDBs of LASSO]{ \label{fig:lsdb}
  \includegraphics[trim=0 0 10pt 24pt, clip,width=.5\textwidth, height = 0.25\textheight] {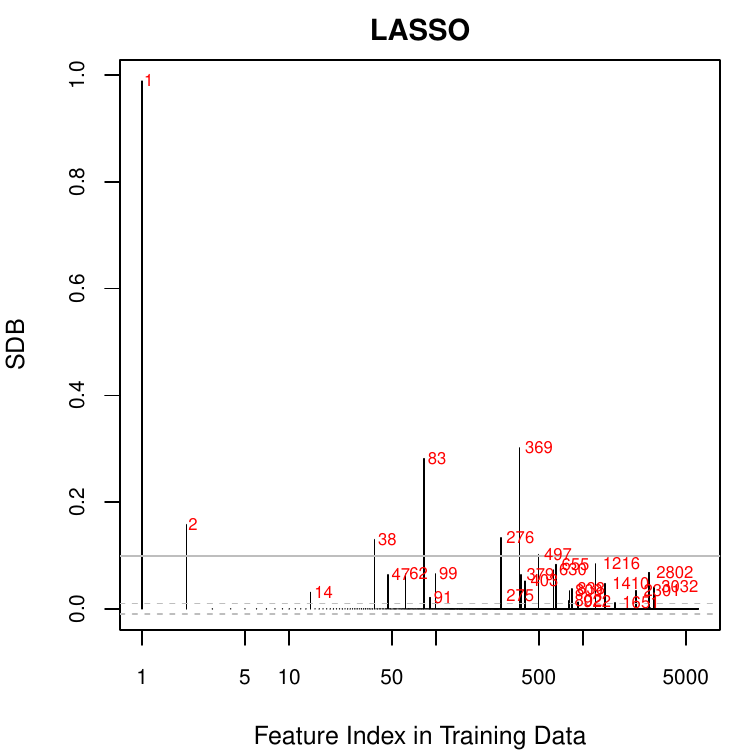} 
}    

\subfloat[SDBs of BLRHL with $t(\df = 1, \log(w) = -10)$]{\label{fig:sdball}
  \includegraphics[trim=0 0 10pt 24pt, clip,width=.5\textwidth, height = 0.25\textheight] {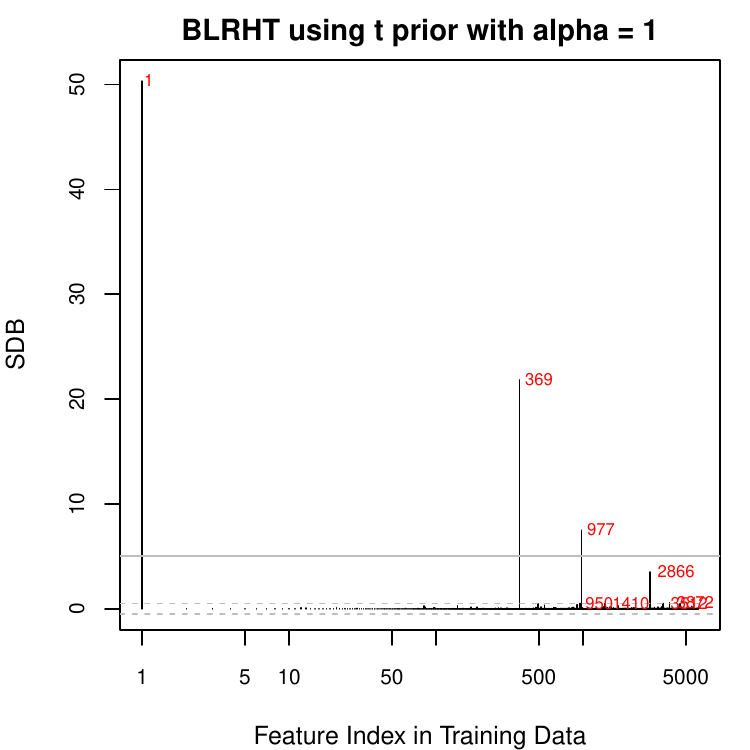}
}
\subfloat[SDBs from rerunning BLRHL for top 50 Genes]{ \label{fig:bsdtop50}
  \includegraphics[trim=0pt 0 10pt 24pt, clip,width=.5\textwidth, height = 0.25\textheight] {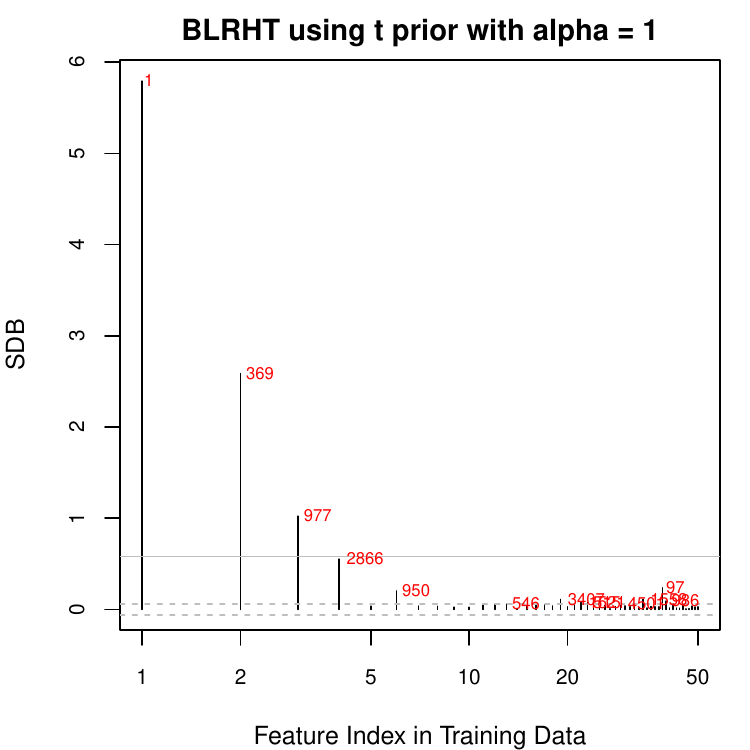}
}

\label{fig:sdbs}
\vspace*{0pt}
\end{figure}

We use leave-one-out cross-validation (LOOCV) to obtain the predictive probabilities for each of the methods considered here. One advantage of LOOCV is that the number of training cases  is only one less than the sample size in the whole dataset; therefore LOOCV predictive measures are believed to be the closest to the out-of-sample predictions based on the whole dataset. Figure \ref{fig:sdbs} shows the SDBs of BLRHL and LASSO when the 2nd case was left out as a test case and the remaining cases were used as training. The 2nd case was chosen to present in this article without any particular reason, as the results are similar for each case left out. P-values given by the F-statistic are calculated for all 102 cases. Figure \ref{fig:bsdtop50} shows the results of rerunning BLRHL using a \textit{t} prior on only the top 50 genes selected using the run with all 6033 genes. The results for BLRHL using the Horseshoe and NEG priors are nearly the same as using the $t$ prior; for this reason they are not shown here. These plots show that BLRHL methods distinctively select fewer than 10 genes by thresholding relative SDBs with 0.01. The F-statistic ranks more than 1000 genes with p-values smaller than 0.01; these genes are actually highly correlated and contain redundant information, therefore they are omitted by BLRHL.  We note that, except for gene 1, all other top genes selected by BLRHL have very low F-statistic ranks (\textit{e.g.} genes 369, 977 and 2866 --- recall that the index is just the F-statistic rank).  We see that LASSO gives many non-zero but small SDBs greater than the value of 0.01 times the maximum SDB, hence LASSO includes many more genes than BLRHL. However, we notice that LASSO omits gene 977, which is ranked the 3rd by BLRHL (later we will use cross-validation to show that this gene is indeed important).

Table \ref{tbl:loocv} shows LOOCV predictive performances measured by AMLP and error rate. In this table we have also included the results of  six other methods reported by \citet{Dettling04}. We see that BLRHL methods are substantially better than many other methods; compared to LASSO, BLRt gains 47\% reduction in AMLP.   

\begin{table}[htp]
\centering
\small
\caption{Comparison of LOOCV predictive performances of BLRHL and others. BLRt, BLRghs and BLRneg are BLRHL using \textit{t}, GHS and NEG priors respectively.}
\vspace*{-5pt}
\begin{tabular}{l|c@{~}c@{~}c@{~}c@{~}c@{~}c@{~}c@{~}c@{~}c@{~}c@{~}}
Methods& BLRt & BLRghs & BLRneg & LASSO & Bagboost & PAM & DLDA & SVM &RanFor &kNN \\[4pt]  
\hline
%\# genes       & 6033 & 6033 & 6033 & 6033  & 200 & 200 & 200 & 200 & 200 & 200 \\[4pt]
AMLP      & .156 & .158 &.152  & .274  & - & - & - & -  & - & -\\[4pt]
ER (\%)   & 6.86 & 7.84 & 7.84 & 10.8   & 7.53 & 16.5 & 14.2 & 7.88 &
9.00 &10.59
\end{tabular}
\label{tbl:loocv}
\vspace*{0pt}
\end{table}

\begin{figure}[t]
\centering
\caption{Comparisons of log predictive probabilities at true class labels.}\label{fig:predprob}
\subfloat[BLRHL vs LASSO]{\label{fig:blpred}
  \includegraphics[width=.45\textwidth, height = 0.25\textheight] {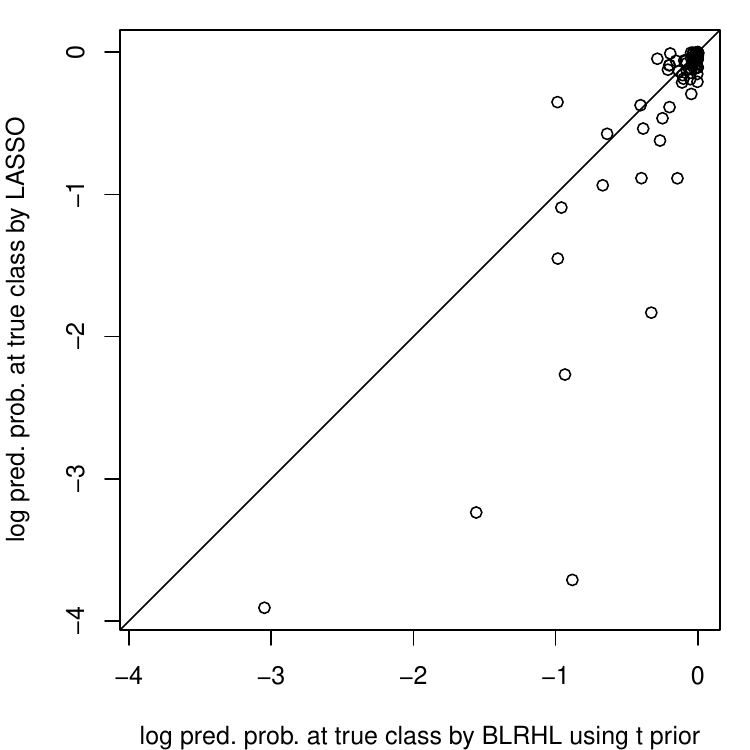}
}
\subfloat[$t$ vs NEG]{ \label{fig:bnpred}
  \includegraphics[width=.45\textwidth, height = 0.25\textheight] {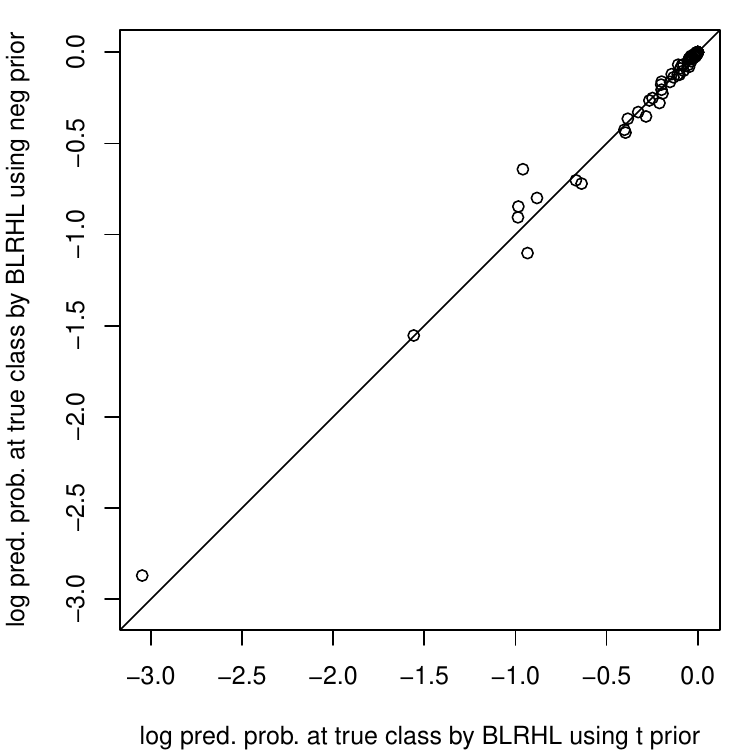}
}
\vspace*{-20pt}

\end{figure}

Figure \ref{fig:predprob} shows the scatterplots of log predictive probabilities at the true class labels for BLRHL and LASSO.  The difference in AMLPs between BLRHL (using a \textit{t} prior) and LASSO is statistically significant, with a p-value of $4.6\times 10^{-4}$ calculated using a paired one-sided \textit{t} test. The performances of BLRHL using \textit{t}/GHS/NEG priors are nearly the same, as shown by Figure \ref{fig:bnpred}. 

\begin{figure}[htp]
\centering
\caption{Scatterplots of some top genes selected by BLRHL. The two red numbers or +  in \ref{fig:3genes}  label the two cases misclassified in LOOCV.} 
\vspace*{-10pt}
\subfloat[]{\label{fig:top12}
\includegraphics[width=.3\textwidth, height=0.2\textheight]{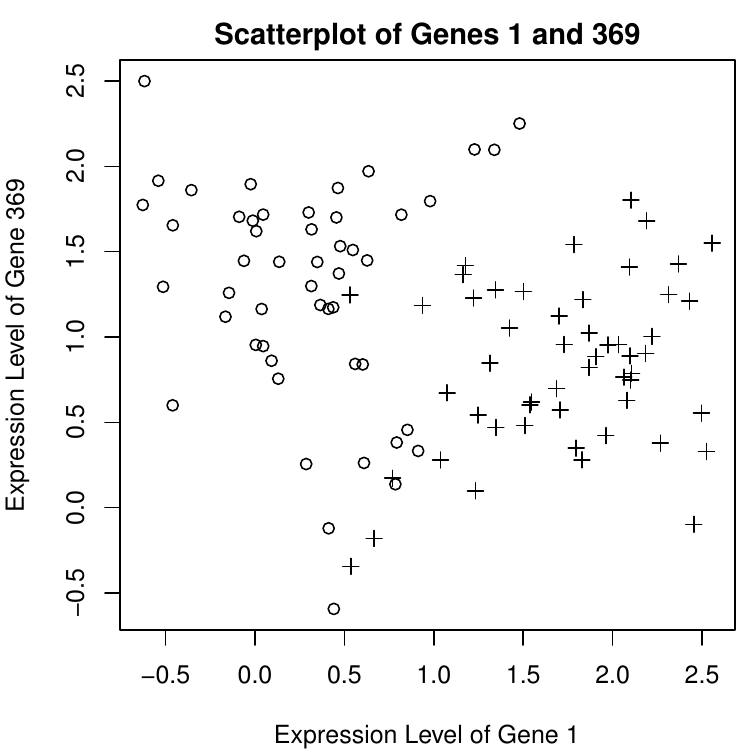}
}       
\subfloat[]{\label{fig:top24}
\includegraphics[width=.3\textwidth, height=0.2\textheight]{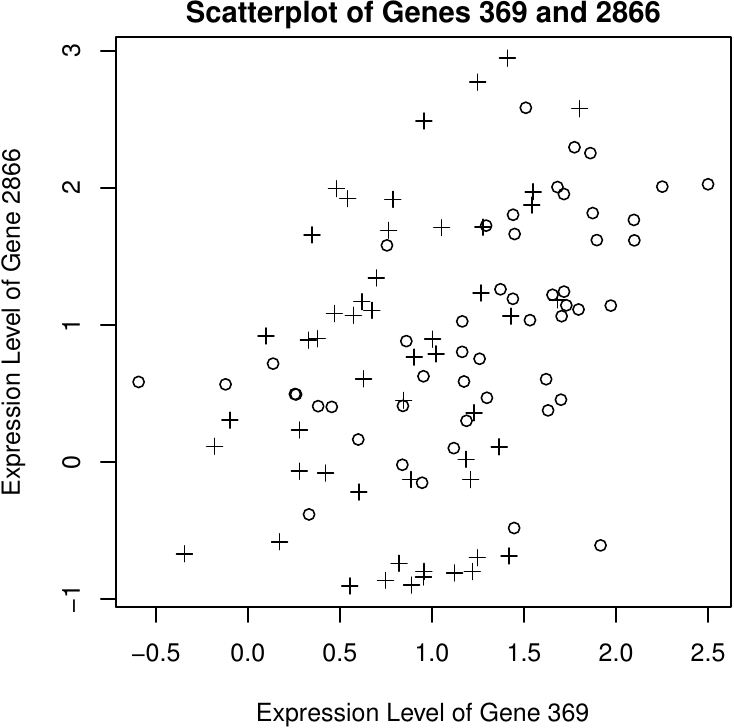}
}
\subfloat[]{\label{fig:3genes}
\includegraphics[width=.3\textwidth, height=0.2\textheight]{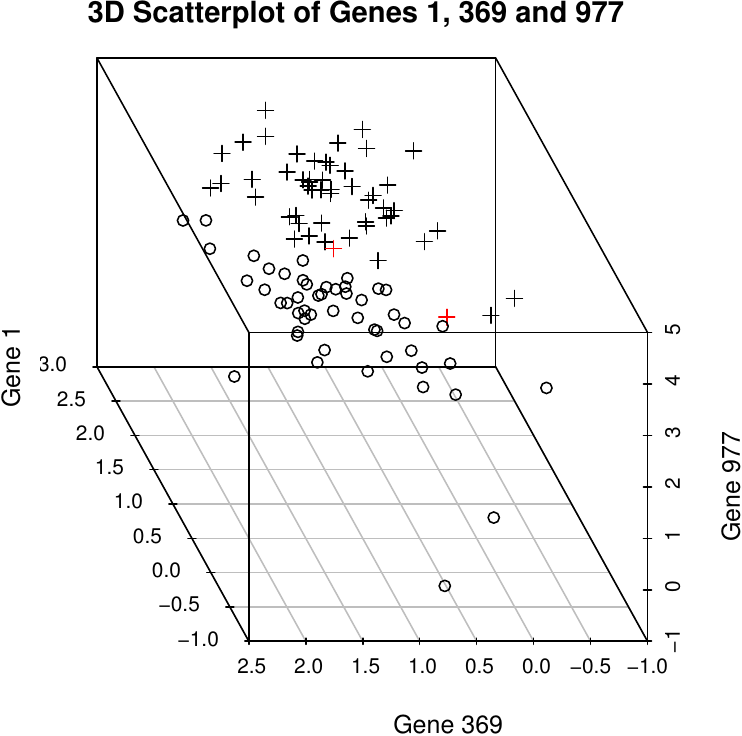}
}
\label{fig:fittingresults}
\vspace*{-20pt}
\end{figure}

To assess the gene selection results of BLRHL, Figure \ref{fig:fittingresults} shows some scatterplots of the top ranking genes (1,369,977,2866).  From these plots we see that genes 369, 977, and 2866 are weakly differentiated across two classes but are useful because they are correlated with the most differentiated gene 1. Gene 2866 is ranked lower because it is correlated with gene 977 as shown by Figure \ref{fig:top24}. Figure \ref{fig:3genes} show that the combination of genes 1, 369 and 977 provides a clear separation for the normal and cancerous tissues.

We further compare the top 3 genes found by BLRHL with other small gene subsets by looking at their LOOCV predictive power. We reran BLRHL (using \textit{t} priors with $\alpha = 1$) on the dataset containing only  genes of a fixed subset (in LOOCV fashion) in order to obtain the predictive power of the given gene subset; these results are shown in Table \ref{tbl:subsets}.  The subset of genes 1, 369 and 977 is substantially better than other subsets in separating the two classes; this is confirmed by the 3D scatterplots (Figure \ref{fig:3genes}) of the top 3 genes.  We think that the subset of genes 1, 369, and 977 is worthy of further biological investigations. From Table \ref{tbl:subsets}, we also see that gene 977 (which is omitted by LASSO) is indeed useful because the subset of genes 1, 369 and 977 has  significantly better predictive power than the subset containing only genes 1 and 369; the AMLP is reduced from 0.232 to 0.05 after including gene 977, with a reduction percentage of $78\%$.  By contrast, the third gene selected by LASSO (gene 83) does not reduce AMLP as much as gene 977.  
\begin{table}[htp]
\centering
\small
\caption{LOOCV predictive performances of various gene subsets.}
\vspace*{-5pt}
\begin{tabular}{l|c@{~~}c@{~~}c@{~~}c@{~~}c@{~~}c@{~~}c@{~~}c@{~~}c@{~~}c@{~~}}
Gene subset& 1, 369, 977 & 1, 369 & 1, 2, 3 & 1, 369, 83  \\[2pt]  
\hline
Selected by & BLRHL & BLRHL and LASSO & F-Statistic & LASSO \\[2pt]
AMLP      & .050 & .232 & .240   & .163 \\[2pt]
ER (\%)   & 1.96 & 8.82 &  9.80  & 7.84 
\end{tabular}
\label{tbl:subsets}
\vspace*{-5pt}
\end{table}

\section{Conclusions and Discussions} \label{sec:end} 

In this article we have introduced an MCMC (fully Bayesian) method for learning severely multi-modal posteriors of logistic regression models based on hyper-LASSO priors (non-convex penalties).   With empirical studies, we have shown that our MCMC algorithm can effectively explore the multi-modal posterior, and hence achieves superior out-of-sample predictive performance and desired hyper-LASSO sparsity for feature selection. Our empirical studies have also demonstrated two important facts about the choice of heaviness and scale of hyper-LASSO priors for logistic regression in datasets with super-sparse signals.  First, the choice of the degrees of freedom that control tail heaviness should be appropriate;  priors with tail heaviness similar to Cauchy appear optimal.  Second,  due to the ``flatness'' in the tails of Cauchy,  the shrinkage of large coefficients is very small (\textit{i.e.} small bias); more importantly, the shrinkage is very robust to the choice of scale, which is a distinctive property of Cauchy priors (compared to Laplace and Gaussian priors). In particular, the choice of $-10$ used in this article for the log scale of Cauchy is expected to work well for a wide range of problems with features standardized to have a standard deviation close to 1, \textit{e.g.} binary indicator variables derived from categorical variables. 

In light of the fact that the posterior distributions based on hyper-LASSO priors are severely multi-modal,  summarizing the feature importance by averaging the coefficients over all modes may not the best choice. In particular, when there is a large group of highly correlated features, many features in the group will be selected when using the means of coefficients.  A more sophisticated method for interpreting the fitting results is to use a clustering algorithm to divide the whole Markov chain samples into subpools, look at the subpools separately, and then deliver a list of succinct feature subsets. If one can split Markov chain iterations as this, it will then be better to use the \textit{median} to obtain an importance index, as it can better shrink the coefficients of totally useless features towards $0$ and correct for the skewness of the posterior.  This will demand further development of methods for interpreting the MCMC samples from a multi-modal posterior.  Another very interesting method is to find a feature subset from the MCMC samples that have the best matching (not the best training predictive power) to the full MCMC samples using the so-called reference/projection approach \citep{goutis1998model,dupuis2003variable,piironen2017comparison}.

%The main drawback of BLRHL is that they are still much slower than many others, such as penalized likelihood methods. Therefore, there is still much room for improving the computational efficiency. The difficulties lie in the high-dimensionality and the existing of multiple posterior modes. A possible solution may be to rotate the original feature space with PCA method, and then apply Gibbs sampling to the posterior distribution of the new coefficients for the transformed feature space with lower dimension. An MCMC simulation for the transformed coefficients may travel across multiple modes with fewer iterations. The crucial step of this solution is to devise a method for sampling the original coefficients conditional on the values of transformed coefficients, a technique similar to what described by \citet{li2008cpb}. 

\onehalfspacing

\smallskip

%\section*{Acknowledgements} 

% .

\singlespacing

\bibliographystyle{asa}

\bibliography{tprobit}

\begin{thebibliography}{37}
\newcommand{\enquote}[1]{``#1''}
\expandafter\ifx\csname natexlab\endcsname\relax\def\natexlab#1{#1}\fi

\bibitem[{Armagan et~al.(2010)Armagan, Dunson, and Lee}]{armagan2010bayesian}
Armagan, A., Dunson, D., and Lee, J. (2010), \enquote{Bayesian generalized
  double Pareto shrinkage,} \textit{Biometrika}.

\bibitem[{Bhattacharya et~al.(2012)Bhattacharya, Pati, Pillai, and
  Dunson}]{bhattacharya2012bayesian}
Bhattacharya, A., Pati, D., Pillai, N.~S., and Dunson, D.~B. (2012),
  \enquote{Bayesian shrinkage,} \textit{{arXiv} preprint {arXiv:1212.6088}}.

\bibitem[{Breheny and Huang(2011)}]{breheny2011coordinate}
Breheny, P. and Huang, J. (2011), \enquote{{Coordinate} {Descent} {Algorithms}
  {For} {Nonconvex} {Penalized} {Regression}, {With} {Applications} {To}
  {Biological} {Feature} {Selection},} \textit{The annals of applied
  statistics}, 5, 232--253, {PMID:} 22081779 {PMCID:} {PMC3212875}.

\bibitem[{Carvalho et~al.(2009)Carvalho, Polson, and
  Scott}]{carvalho2009handling}
Carvalho, C.~M., Polson, N.~G., and Scott, J.~G. (2009), \enquote{{Handling
  sparsity via the horseshoe},} \textit{Journal of Machine Learning Research},
  5.

\bibitem[{Carvalho et~al.(2010)Carvalho, Polson, and
  Scott}]{carvalho2010horseshoe}
--- (2010), \enquote{{The horseshoe estimator for sparse signals},}
  \textit{Biometrika}, 97, 465.

\bibitem[{Clarke et~al.(2008)Clarke, Ressom, Wang, Xuan, Liu, Gehan, and
  Wang}]{clarke2008properties}
Clarke, R., Ressom, H.~W., Wang, A., Xuan, J., Liu, M.~C., Gehan, E.~A., and
  Wang, Y. (2008), \enquote{The properties of high-dimensional data spaces:
  implications for exploring gene and protein expression data,} \textit{Nat.
  Rev. Cancer}, 8, 37--49.

\bibitem[{Dettling(2004)}]{Dettling04}
Dettling, M. (2004), \enquote{{BagBoosting for tumor classification with gene
  expression data},} \textit{Bioinformatics}, 20, 3583--3593.

\bibitem[{Dudoit et~al.(2002)Dudoit, Fridlyand, and
  Speed}]{dudoit2002comparison}
Dudoit, S., Fridlyand, J., and Speed, T.~P. (2002), \enquote{{Comparison of
  discrimination methods for the classification of tumors using gene expression
  data},} \textit{Journal of the American Statistical Association}, 97, 77--87.

\bibitem[{Dupuis and Robert(2003)}]{dupuis2003variable}
Dupuis, J.~A. and Robert, C.~P. (2003), \enquote{Variable selection in
  qualitative models via an entropic explanatory power,} \textit{Journal of
  Statistical Planning and Inference}, 111, 77--94.

\bibitem[{Fan and Li(2001)}]{fan2001variable}
Fan, J. and Li, R. (2001), \enquote{Variable Selection via Nonconcave Penalized
  Likelihood and its Oracle Properties,} \textit{Journal of the American
  Statistical Association}, 96, 1348--1360.

\bibitem[{Gelman(2006)}]{gelman2006prior}
Gelman, A. (2006), \enquote{{Prior distributions for variance parameters in
  hierarchical models},} \textit{Bayesian analysis}, 1, 515--533.

\bibitem[{Gelman et~al.(2008)Gelman, Jakulin, Pittau, and
  Su}]{gelman2008weakly}
Gelman, A., Jakulin, A., Pittau, M.~G., and Su, Y. (2008), \enquote{A weakly
  informative default prior distribution for logistic and other regression
  models,} \textit{The Annals of Applied Statistics}, 2, 1360--1383.

\bibitem[{Gilks and Wild(1992)}]{gilks1992adaptive}
Gilks, W.~R. and Wild, P. (1992), \enquote{{Adaptive rejection sampling for
  Gibbs sampling},} \textit{Applied Statistics}, 41, 337--348.

\bibitem[{Goutis and Robert(1998)}]{goutis1998model}
Goutis, C. and Robert, C.~P. (1998), \enquote{Model choice in generalised
  linear models: A Bayesian approach via {Kullback-Leibler} projections,}
  \textit{Biometrika}, 85, 29--37.

\bibitem[{Griffin and Brown(2011)}]{griffin2011bayesian}
Griffin, J.~E. and Brown, P.~J. (2011), \enquote{Bayesian {Hyper-Lassos} with
  {Non-Convex} Penalization,} \textit{Australian \& New Zealand Journal of
  Statistics}, 53, 423--442.

\bibitem[{Kotz and Nadarajah(2004)}]{kotz2004multivariate}
Kotz, S. and Nadarajah, S. (2004), \textit{{Multivariate t distributions and
  their applications}}, Cambridge Univ Pr.

\bibitem[{Kyung et~al.(2010)Kyung, Gill, Ghosh, and
  Casella}]{kyung2010penalized}
Kyung, M., Gill, J., Ghosh, M., and Casella, G. (2010), \enquote{{Penalized
  regression, standard errors, and bayesian lassos},} \textit{Bayesian
  Analysis}, 5, 369--412.

\bibitem[{Li(2012)}]{libcbcsf}
Li, L. (2012), \enquote{{Bias-Corrected Hierarchical Bayesian Classification
  With a Selected Subset of High-Dimensional Features},} \textit{Journal of the
  American Statistical Association}, 107, 120--134.

\bibitem[{Ma et~al.(2007)Ma, Song, and Huang}]{ma2007supervised}
Ma, S., Song, X., and Huang, J. (2007), \enquote{{Supervised group Lasso with
  applications to microarray data analysis},} \textit{BMC Bioinformatics}, 8,
  60.

\bibitem[{Nalenz and Villani(2017)}]{nalenz2017tree}
Nalenz, M. and Villani, M. (2017), \enquote{Tree Ensembles with Rule Structured
  Horseshoe Regularization,} \textit{{arXiv:1702.05008} [stat]}, {arXiv:}
  1702.05008.

\bibitem[{Neal(2010)}]{neal2010mcmc}
Neal, R.~M. (2010), \enquote{{MCMC using Hamiltonian dynamics},} in
  \textit{Handbook of Markov Chain Monte Carlo} (eds S. Brooks, A. Gelman, G.
  Jones, XL Meng). Chapman and Hall/CRC Press.

\bibitem[{Piironen and Vehtari(2016)}]{piironen2016hyperprior}
Piironen, J. and Vehtari, A. (2016), \enquote{On the Hyperprior Choice for the
  Global Shrinkage Parameter in the Horseshoe Prior,}
  \textit{{arXiv:1610.05559} [stat]}, {arXiv:} 1610.05559.

\bibitem[{Piironen and Vehtari(2017)}]{piironen2017comparison}
--- (2017), \enquote{Comparison of Bayesian predictive methods for model
  selection,} \textit{Statistics and Computing}, 27, 711--735.

\bibitem[{Polson and Scott(2010)}]{polson2010shrink}
Polson, N.~G. and Scott, J.~G. (2010), \enquote{Shrink globally, act locally:
  Sparse Bayesian regularization and prediction,} \textit{Bayesian Statistics},
  9, 501--538.

\bibitem[{Polson and Scott(2012{\natexlab{a}})}]{polson2012good}
--- (2012{\natexlab{a}}), \enquote{Good, great, or lucky? Screening for firms
  with sustained superior performance using heavy-tailed priors,} \textit{The
  Annals of Applied Statistics}, 6, 161--185.

\bibitem[{Polson and Scott(2012{\natexlab{b}})}]{polson2012local}
--- (2012{\natexlab{b}}), \enquote{Local shrinkage rules, Levy processes and
  regularized regression,} \textit{Journal of the Royal Statistical Society:
  Series B {(Statistical} Methodology)}, 74, 287--311.

\bibitem[{Polson and Scott(2012{\natexlab{c}})}]{polson2012half-cauchy}
--- (2012{\natexlab{c}}), \enquote{On the {half-Cauchy} prior for a global
  scale parameter,} \textit{Bayesian Analysis}, 7, 887--902.

\bibitem[{Singh et~al.(2002)Singh, Febbo, Ross, Jackson, Manola, Ladd, Tamayo,
  Renshaw, D'Amico, Richie, and Others}]{singh2002gene}
Singh, D., Febbo, P.~G., Ross, K., Jackson, D.~G., Manola, J., Ladd, C.,
  Tamayo, P., Renshaw, A.~A., D'Amico, A.~V., Richie, J.~P., and Others (2002),
  \enquote{{Gene expression correlates of clinical prostate cancer behavior},}
  \textit{Cancer cell}, 1, 203--209.

\bibitem[{Tibshirani(1996)}]{tibshirani1996regression}
Tibshirani, R. (1996), \enquote{{Regression Shrinkage and Selection via the
  Lasso},} \textit{Journal of the Royal Statistical Society: Series B
  (Methodological)}, 58, 267--288.

\bibitem[{Tibshirani et~al.(2002)Tibshirani, Hastie, Narasimhan, and
  Chu}]{tibshirani2002dmc}
Tibshirani, R., Hastie, T., Narasimhan, B., and Chu, G. (2002),
  \enquote{{Diagnosis of multiple cancer types by shrunken centroids of gene
  expression},} \textit{Proceedings of the National Academy of Sciences}, 99,
  6567.

\bibitem[{Tolosi and Lengauer(2011{\natexlab{a}})}]{tolosilengauer2011}
Tolosi, L. and Lengauer, T. (2011{\natexlab{a}}), \enquote{{Classification with
  correlated features: unreliability of feature ranking and solution},}
  \textit{Bioinformatics}, 27, 1986--1994.

\bibitem[{Tolosi and Lengauer(2011{\natexlab{b}})}]{tolosi2011classification}
--- (2011{\natexlab{b}}), \enquote{Classification with correlated features:
  unreliability of feature ranking and solutions,} \textit{Bioinformatics}, 27,
  1986--1994.

\bibitem[{van~der Pas et~al.(2014)van~der Pas, Kleijn, and van~der
  Vaart}]{van_der_pas2014horseshoe}
van~der Pas, S.~L., Kleijn, B. J.~K., and van~der Vaart, A.~W. (2014),
  \enquote{The Horseshoe Estimator: Posterior Concentration around Nearly Black
  Vectors,} \textit{{arXiv:1404.0202} [math, stat]}.

\bibitem[{Wang et~al.(2014)Wang, Liu, and Zhang}]{wang2014optimal}
Wang, Z., Liu, H., and Zhang, T. (2014), \enquote{Optimal computational and
  statistical rates of convergence for sparse nonconvex learning problems,}
  \textit{Annals of statistics}, 42, 2164.

\bibitem[{Yi and Ma(2012)}]{yi2012hierarchical}
Yi, N. and Ma, S. (2012), \enquote{Hierarchical Shrinkage Priors and Model
  Fitting for High-dimensional Generalized Linear Models,} \textit{Statistical
  applications in genetics and molecular biology}, 11, {PMID:} 23192052
  {PMCID:} {PMC3658361}.

\bibitem[{Zhang(2010)}]{zhang2010nearly}
Zhang, C. (2010), \enquote{Nearly unbiased variable selection under minimax
  concave penalty,} \textit{The Annals of Statistics}, 38, 894--942, {MR:}
  {MR2604701} Zbl: 05686523.

\bibitem[{Zou(2006)}]{zou2006adaptive}
Zou, H. (2006), \enquote{The Adaptive Lasso and Its Oracle Properties,}
  \textit{Journal of the American Statistical Association}, 101, 1418--1429.

\end{thebibliography}

% !TEX root =  bplrpaper.tex
% control space around section
\titlespacing*{\section}{0pt}{-5pt}{-2pt}
\titlespacing*{\subsection}{0pt}{-5pt}{-2pt}
\appendix

\bigskip
\onehalfspacing

\section*{Appendices}

\section{Computational Method for BLRHL}
\label{sec:appendix}

This section is a  continued discussion from Section \ref{sec:gibbs} about our
computational method.

\subsection{Initial Values for Gibbs Sampling}

The initial values for $\delta_{0:p, 1:K}$ are coefficients of the Bayes
discriminant rule based on Gaussian distributions whose mean vectors are
estimated by the medians of Markov chain samples produced by the method
described in \cite{libcbcsf} and whose covariance matrix is estimated by
an equally weighted average of the sample covariance and the identity matrix.

\subsection{Updating $\delta_{0:p,1:K}$ with Hamiltonian Monte
Carlo}\label{sec:hmc}

Suppose that we want to sample from a $d$-dimensional distribution with PDF
proportional to $\exp (- U(\mb q))$ , or construct a transformation leaving it
invariant.  For our problem, $U(\mb q)$ is the minus log of the posterior distribution
of $\mb q = \mb \delta_{0:p, 1:K}$ (\textit{i.e.} the minus log of \eqref{eqn:postdelta}).

We will augment $\mb q$ with a set of auxiliary variables $\mb p$ that are independently
distributed with $N(0,1)$ and are independent of $\mb q$. For this purpose we will
randomly draw a $\mb p$ independently from $N(0,1)$. In physics, $\mb p$ is
interpreted as momentum’s of particles. Next we will transform $(\mb q,\mb p)$
in a way that leaves invariant $\exp (H(\mb q, \mb p))$  --- the joint
distribution of $(\mb q,\mb p)$, where $H(\mb q, \mb p)$ is often called
\textbf{Hamiltonian}, which is given by:
  \begin{equation*}
  H(\mb q,\mb p) = U(\mb q) + K(\mb p) = U(\mb q) + \frac{1}{2} \sum_{i=1}^d
p_i^2.
  \end{equation*}
At the end of this transformation, we will discard $\mb q$, obtaining a new $\mb
p$ that is still distributed with $\exp(-U(\mb p))$.

The method for transforming $(\mb q, \mb p)$ is inspired by Hamiltonian
dynamics, in which $(\mb q,\mb p)$ moves along a continuous time $\tau$
according to the following differential equations:  
\begin{eqnarray*}
   \frac{d q_i(\tau)}{d\tau} &=& \frac{\partial H}{\partial p_i} \ \ \ = \ \ \ 
\frac{\partial K}{\partial p_i}\ \ \   = \ \ \  p_i\\
   \frac{d p_i(\tau)}{d\tau} &=& - \frac{\partial H}{\partial q_i} \ \ \  = \ \
\  - \frac{\partial U}{\partial q_i}
  \end{eqnarray*}
It can be shown that this Hamiltonian dynamic keeps $H$ unchanged and preserves
volume (see details from \cite{neal2010mcmc}). These are the crucial properties
of Hamiltonian dynamics that make it a good proposal distribution for Metropolis
sampling. 

In computer implementation, Hamiltonian dynamics must be approximated by
discretized time, using small stepsize $\epsilon$. Leapfrog transformation is
one of such methods, which is shown to be better than several other
alternatives. \textit{One} leapfrog transformation with stepsize $\epsilon_i$ is
described as follows: 

  \smallskip

  \centerline{\framebox{\bf One  Leapfrog Transformation}}
  \vspace*{-20pt}
  \begin{eqnarray*}
   p_i &\leftarrow& p_i - (\epsilon_i/2) \ \frac{\partial U}{\partial q_i}
(q_i),\\
   q_i &\leftarrow& q_i + \epsilon_i\ p_i,\\
   p_i &\leftarrow& p_i - (\epsilon_i/2) \ \frac{\partial U}{\partial q_i}
(q_i).
  \end{eqnarray*}
Note that we apply leapfrog transformations independently to each pair
$(q_i,p_i)$ using different stepsizes. By applying a series of leapfrog
transformations, we \textit{deterministically} transform $(q_i,p_i)$ to a new
state, denoted by $(q_i^*, p_i^*)$, for $i=1,\ldots, d$. This transformation has
the following properties:
 \begin{itemize}
   \item The value of $H$ is nearly unchanged if $\epsilon_i$ is small enough. This is because each leapfrog transformation is a good approximation
    to Hamiltonian dynamics.
   \item Reversibility:  following the same series of leapfrog transformations,
   $(q_i^*, -p_i^*)$ will be transformed back to $(q_i, -p_i)$. We therefore add
   a negation ahead of these leapfrog transformations to form an exactly
  ``reversible'' transformation between $(q_i, -p_i)$ and $(q_i^*, p_i^*)$.
   \item Volume preservation: the Jacobian of this transformation is 1.
  \end{itemize}

A series of leapfrog transformations cannot leave $H$ exactly unchanged, but we
will use it only as a proposal distribution in Metropolis sampling. That is, at
the end of the leapfrog transformations, $(\mb q^*, \mb p^*)$ will be accepted
or rejected randomly according to Metropolis acceptance probability. As a
summary, the algorithm of Hamiltonian Monte Carlo is presented completely below:

\smallskip 

\centerline{\framebox{\bf\ Hamiltonian Monte Carlo (HMC) with Leapfrog
Transformations}}

{\sf

Starting from current state $\mb q$, update it with the following steps:

\begin{quote}

\textbf{Step 1:} Draw elements of $-\mb p$ independently from $N(0,1)$

\textbf{Step 2:} Transform $(\mb q, -\mb p)$ with the following two steps:

 \textbf{(a)} Negate $-\mb p$ to $\mb p$.

 \textbf{(b)} Apply the leapfrog transformation $\ell$ times to transform $(\mb q,
\mb p)$ to a new state $(\mb q^*, \mb p^*)$. A trajectory connecting the states
along these $\ell$  transformations is called the \textit{leapfrog trajectory} with
length $\ell$.

\hangindent=\parindent \textbf{Step 3:} Decide whether or not to accept $(\mb
q^*, \mb p^*)$  with a probability given by:
\begin{equation}
\min\left(1,\ \ \exp\left( - \Big[H(\mb q^*,\mb p^*) - H(\mb q, - \mb p)\Big]\
\right) \ \right). \label{eqn:acceptprob}
\end{equation}
 If the result is a rejection, set $ (\mb q^*, \mb p^*) = (\mb q, -\mb p)$.

\end{quote}

At last, retaining $\mb q^*$, with $\mb p^*$ discarded.
}

To implement HMC, we need to choose appropriate stepsizes $\epsilon_i$ and
$\ell$ --- the length of the leapfrog trajectory; these stepsizes determine how well the
leapfrog transformation can approximate Hamiltonian dynamics. If $\epsilon_i$ is
too large, the leapfrog transformation may diverge, resulting in a very high rejection
rate and very poor performance; otherwise, it may move too slowly, even though
there is a very low rejection rate. An ad-hoc choice is a value close to the
reciprocal of the square root of the 2nd-order partial derivative of $U$ with
respect to $q_i$, which automatically accounts for the width of the posterior
distribution of $q_i$. We therefore adjust the reciprocals by an adjustment
factor $\epsilon$ (which usually should be between 0.1 and 0.5, called the
\textit{HMC stepsize adjustment}. The exact value of the adjustment factor can
be chosen empirically such that the HMC rejection rate is less than but close to
$0.2$; this is because there is often a critical point beyond which the Hamiltonian diverges. A
value slightly smaller than this critical point often works the best; according
to our experiences, a value close to $0.25$ often works well. A good thing about
this choice is that it is independent of the choice of $\ell$ --- the length of the
leapfrog trajectory, because the value of the Hamiltonian (actually the whole
leapfrog trajectory) changes nearly cyclically as long as it doesn't diverge.

After we determine the stepsize adjustment $\epsilon$, we will determine the
length of the leapfrog trajectory $\ell$.  The fact is that the appropriate values
of $\ell$ are different in two phases. In the \textit{initial phase}, a small
value $\ell_1$ should be used such that Gibbs sampling quickly
dissipates the value of $U$ and more frequently updates the
hyperparameter $w$. The exact choice of $\ell$ for the initial phase can be made
empirically by looking at how fast Gibbs sampling converges with different
values of $\ell$. For our problem, $\ell_1 = 10$ or 5 seems to work well. Another reason for the initial phase is that a very long trajectory
starting from the initial value has very high chance of being rejected. After
running the initial phase for a while, we need to choose a larger value (denoted by
$\ell_2$) to suppress random walk; this phase is called the \textit{sampling phase}.
The advantage of using HMC instead of other samplers is that HMC can keep moving
in the direction determined by the gradients of $U$ without random walk. We
therefore should choose fairly large $\ell$ (at least larger than 1; when
$\ell=1$, HMC is equivalent to the Langevin Metropolis-Hasting method) such that the leapfrog
transformation can reach a distant point from the starting one. However, if
$\ell$ is excessively large, the leapfrog transformation will reverse the
direction and move back to the region near the starting point. The choice of
$\ell$ for this phase can be made empirically by looking at the curve of
the distance of $\mb q$ from the origin along a very long trajectory, which changes
cyclically. We will choose the largest $\ell$ such that leapfrog transformation
can move in a direction, \textit{i.e.} the distance of $\mb q$ changes monotonically. From
our experiences, $\ell_2 = 50$ or 100 works well for many problems.

To apply HMC to sample \eqref{eqn:postdelta}, we need to compute the 1st-order partial derivatives of $-\log(P(\mb\delta_{0:p,1:K}|\mb\sigma_{0:p}^2))$ with respect to $\delta_{jk}$ for $j = 0,\ldots, p, k = 1,\ldots, K$, which is equal to the sum of the following two partial derivatives of $L$ and minus log prior:
 \begin{eqnarray}
  - \dfrac{\partial \log (L(\mb\delta_{0:p,1:K}))}{\partial \delta_{jk}}
  & = & \sum_{i=1}^n x_{ij}(P(y_i=k+1|x_{ij}, \mb\delta_{0:p, 1:K})) - I
(y_i=k+1)),\\
  - \dfrac{\partial \log(P(\mb \delta_{0:p, 1:K}|\mb \sigma^2_{0:p}))}{\partial
\delta_{jk}}
  & = & \left(\delta_{jk} - \sum_{k=1}^K \delta_{jk}/C\right)\Big / \sigma_j^2.
 \end{eqnarray}

An ad-hoc choice of the stepsizes $\epsilon_{jk}$ is a value close to the reciprocal of the 2nd-order derivatives of $U$. We also use an estimate of the 2nd-order derivatives of $U$, which should be independent of current values of $\mb\delta_{0:p, 1:K}$ but could be dependent on $\mb\sigma_{0:p}^2$. They are the sum of the following two values:
\begin{eqnarray*}
  - \dfrac{\partial^2 \log (L(\mb\delta_{0:p, 1:K}))}{\partial^2 \delta_{jk}}
 & \approx & \sum_{i=1}^n x_{ij}^2/4,\\
 - \dfrac{\partial^2 \log(P(\mb \delta_{0:p,1:K}|\mb
\sigma_{0:p}^2))}{\partial^2 \delta_{jk}}
 & = & \frac{C-1}{C}\frac{1}{\sigma_j^2}.
\end{eqnarray*}

\subsection{Restricted Gibbs Sampling}\label{sec:rgs}

When $p$ is large, the dominating computing in applying HMC is obtaining values of the linear functions $\delta_{0,k} + \mb x_{i,1:p} \mb\delta_{1:p,k}$ for $i=1,\ldots, n$ and $k=1,\ldots,K$, with which we can compute the log likelihood and its partial derivatives with respect to $\mb\delta_{0:p, 1:K}$ very easily. 

A belief in high-dimensional classification is that most features are irrelevant and therefore most coefficients concentrate very close to 0 in a local mode of the posterior; it is therefore useless to update them very often. A useful computational trick for reducing computation time is that, for each iteration of Gibbs sampling, we update only those features with $\sigma_j$ greater than a small threshold $\zeta$ without much loss of efficiency. However, even a fairly small $\zeta$  can cut off many coefficients from being updated. The consequence of this is that the computation time for each iteration of Gibbs sampling is reduced substantially, since we can reuse from the last iteration the sum of a large number of $x_{i,j}\delta_{j,c}$ related to those small coefficients, which are to be fixed. We call this trick \textbf{restricted Gibbs sampling}. We want to point out that this method can be justified with Markov chain theory, therefore our computation using this trick is still an exact Markov chain simulation. The essential effect of this trick is updating those important features more often than a large number of coefficients for irrelevant features. However, note that when $\zeta$ is chosen to be very large, only a few (say ones) coefficients are updated using HMC, and we therefore lose the ability of HMC in suppressing random walk. The consequence of this is that the Markov chain may have more difficulty travelling across the modes, as travelling from one mode to another requires a very small coefficient to be updated to a large one. Further research is needed to find the optimal choice of $\zeta$. The implementation used in our examples chose $\zeta=0.05$ when $\alpha$ is set to $1$, for which about $10\%$ of coefficients are updated in each iteration.

\section{Notations of Prior and MCMC Settings in  BLRHL}\label{sec:settings}

First, one needs to choose the prior type from \texttt{t}, \texttt{ghs}, and \texttt{neg}. For each choice of prior type, one needs to set these parameters for the prior and MCMC computation: 

\begin{itemize}[topsep=0pt, partopsep=0pt, itemsep=0pt]
\item $\alpha$, $\log(w)$: degree freedom (df) and log square scale of \texttt{t}/\texttt{ghs}/\texttt{neg} prior.

\item $n_1,\ell_1$: number of Gibbs sampling iterations and length of trajectory in initial phase.

\item $n_2,\ell_2$: number of Gibbs sampling iterations and length of trajectory in sampling phase.

\item $\zeta$: the coefficients with $\sigma_j$ smaller than $\zeta$ are fixed in current HMC updating.

\item $\epsilon$: stepsize adjustment multiplied to the 2nd order partial derivatives of log posterior.

\item the prior variances $\sigma_0^2$ for the intercepts are always set to 2000. 

\end{itemize}

\section{R Code}
An R package under development with manual and a demonstration R file using the example described in Section \ref{sec:sim2000} and \ref{sec:500datasets} are available from the publisher's and the first author's website.

\end{document}